%% file: paper.tex
\documentclass[runningheads]{llncs}
\usepackage[T1]{fontenc}
\usepackage{graphicx}
\usepackage[utf8]{inputenc}
\usepackage[title]{appendix}
\usepackage[english]{babel}
\usepackage{blindtext}

\usepackage{latexsym}      
\usepackage{mathtools}	   
\usepackage{xspace}   
\usepackage{booktabs}
\usepackage{epsfig}
\usepackage{xcolor}   
\usepackage{tabularx} 
\usepackage{pifont}
\usepackage[font=small,labelfont=bf]{caption}
\usepackage{balance}
\usepackage{multirow}
\usepackage{graphicx}
\usepackage{subcaption}
\usepackage{enumitem}
\usepackage{comment}
\usepackage{listings}
\usepackage{multirow}
\usepackage{url}

\usepackage{cleveref}
\crefformat{section}{#2\S#1#3}
\Crefformat{section}{#2Section~#1#3}
\crefname{section}{\S}{\S\S}
\Crefname{section}{Section}{Sections}
\crefname{equation}{Equation}{Equations}
\Crefname{equation}{Equation}{Equations}
\crefname{table}{Table}{Tables}
\Crefname{table}{Table}{Tables}
\crefname{figure}{Fig.}{Figs.}
\Crefname{figure}{Figure}{Figures}
\crefname{algorithm}{Algorithm}{Algorithms}

\begin{document}
%
\title{Characterizing the Networks Sending Enterprise Phishing Emails}
%
%
\author{Elisa Luo\inst{1,2,3}\and
Liane Young\inst{1}\and
Grant Ho\inst{3,4}\and M. H. Afifi\inst{2} \and Marco Schweighauser\inst{2}\and Ethan Katz-Bassett\inst{1}\and Asaf Cidon\inst{1,2}}
%
\authorrunning{E. Luo et al.}
\institute{Columbia University\and Barracuda Networks\and UC San Diego\and University of Chicago}
%
%

\input{macros}

\maketitle              
\input{abstract}
\input{intro}
\input{background}
\input{dataset}
\input{results}
\input{case_studies}
\input{detector}
\input{discussion}

\begin{credits}
\subsubsection{\ackname} We thank our anonymous reviewers and shepherd for their insightful and constructive feedback. This work was supported in part by the NSF CNS-2106530, CNS-2143868 and CNS-2403431 awards, and the UCSD CSE Postdoctoral Fellows program.

\end{credits}
\clearpage

%
%
%
%

\sloppy
\bibliographystyle{splncs04}
\bibliography{references,references-asaf}

\clearpage
\input{appendix}

\end{document}

%% file: macros.tex
\newcommand{\cudaFullName}{{Barracuda Networks}\xspace}
\newcommand{\cuda}{{Barracuda}\xspace}

\newcommand{\ttext}[1]{{\sf #1}}
\newcommand{\unclear}[1]{{\textcolor{red}{\textit{#1}}}}
\newcommand{\cut}[1]{\textcolor{blue}{#1}}  
\newcommand{\new}[1]{\textcolor{violet}{#1}}
\newcommand{\ecut}[1]{\textcolor{purple}{#1}}
\newcommand{\enew}[1]{\textcolor{black}{#1}}

\newcommand{\todo}[1]{\textcolor{red}{TODO: #1}}
\newcommand{\revision}[1]{\textcolor{violet}{#1}}
\newcommand{\tbd}[1]{\todo{#1}}
\newcommand{\grant}[1]{\textcolor{red}{\noindent[Grant: #1]}}
\newcommand{\ekb}[1]{\textcolor{purple}{\noindent[Ethan: #1]}}
\newcommand{\afifi}[1]{\textcolor{blue}{\noindent[Afifi: #1]}}
\newcommand{\elisa}[1]{\textcolor{magenta}{\noindent[Elisa: #1]}}
\newcommand{\liane}[1]{\textcolor{cyan}{\noindent[Liane: #1]}}
\newcommand{\Yifan}[1]{\textcolor{red}{\noindent[Yifan: #1]}}

\newcommand{\eg}{e.g.,\xspace}
\newcommand{\ie}{i.e.,\xspace}
\newcommand{\from}{\textsc{from\_email}\xspace}

\newcommand{\trainingMonths}{April--June 2018\xspace}
\newcommand{\evalMonths}{July--October 2018\xspace}
\newcommand{\numMonths}{7\xspace}

\newcommand{\noForgePercent}{85\%\xspace}
\newcommand{\totalOrgs}{92\xspace}
\newcommand{\numEmails}{113,083,695\xspace}

\newcommand{\numPhishEmails}{1,902\xspace}
\newcommand{\numIncidents}{180\xspace}

\newenvironment{denseitemize}{
\begin{itemize}[topsep=2pt, partopsep=0pt, leftmargin=1.5em]
  \setlength{\itemsep}{2pt}
  \setlength{\parskip}{0pt}
  \setlength{\parsep}{0pt}
}{\end{itemize}}

\newenvironment{denseenum}{
\begin{enumerate}[topsep=2pt, partopsep=0pt, leftmargin=1.5em]
  \setlength{\itemsep}{2pt}
  \setlength{\parskip}{0pt}
  \setlength{\parsep}{0pt}
}{\end{enumerate}}

\def\UrlBreaks{\do\/\do-}

%% file: abstract.tex
\begin{abstract}
\vspace{-3ex}
Phishing attacks on enterprise employees present one of the most costly and potent threats to organizations. We explore an understudied facet of enterprise phishing attacks: the email relay infrastructure behind successfully delivered phishing emails. We draw on a dataset spanning one year across thousands of enterprises, billions of emails, and over 800,000 delivered phishing attacks. Our work sheds light on the network origins of phishing emails received by real-world enterprises, differences in email traffic we observe from networks sending phishing emails, and how these characteristics change over time.

Surprisingly, we find that over one-third of the phishing email in our dataset originates from highly reputable networks, including Amazon and Microsoft. Their total volume of phishing email is consistently high across multiple months in our dataset, even though the overwhelming majority of email sent by these networks is benign. In contrast, we observe that a large portion of phishing emails originate from networks where the vast majority of emails they send are phishing, but their email traffic is not consistent over time. Taken together, our results explain why no singular defense strategy, such as static blocklists (which are commonly used in email security filters deployed by organizations in our dataset), is effective at blocking enterprise phishing. Based on our offline analysis, we partnered with a large email security company to deploy a classifier that uses dynamically updated network-based features. In a production environment over a period of 4.5 months, our new detector was able to identify 3-5\% more enterprise email attacks that were previously undetected by the company's existing classifiers. 

\end{abstract}

%% file: intro.tex
\section{Introduction}
Phishing attacks remain one of the most costly threats to enterprises, resulting in billions of dollars in losses~\cite{fbi-26b}, disrupting critical infrastructure~\cite{colonial},
and imperiling national security~\cite{nuclear-phish,presidentialEmailLeaks}.
Although a substantial body of research has proposed various countermeasures~\cite{FIRE,simoiu2020targeted,databreaches,allodi2019need,khonji2013phishing,medvet2008visual}, the continued and widespread success of phishing illustrates the need for better defenses.


\enew{Prior work on spam and mass phishing campaigns have proposed simple blocklists as a defense~\cite{FIRE,qian2010network,hao2009detecting}, with an implicit premise that phishing emails will emanate from stable and predominantly malicious servers. 
In practice, many organizations do use IP address-based blocklists, including many of the ones in our email dataset (\cref{sec:no-filter}).
However, these methods have limited efficacy against enterprise phishing; in our dataset alone, they fail to block hundreds of thousands of phishing emails.
This limitation arises because blocklists often grow stale and can suffer from slow updating delays as they tend to rely on user reports~\cite{sun2024victims}. 
Given the evolving web and cloud landscape -- where IP addresses and servers frequently change ownership, host multiple independent tenants in parallel, and/or host legitimate services that have been compromised by attackers -- the efficacy of such methods against modern enterprise phishing remains unclear.}

Our work seeks to better understand the network characteristics of modern enterprise phishing attacks, with the aim of identifying more nuanced and effective defenses that go beyond static blocklisting.
Performing this analysis, which explores and characterizes the email infrastructure responsible for routing phishing emails, requires a large-scale dataset of both phishing and clean emails from many enterprises.
Although some studies have involved large corpora of phishing emails~\cite{lateral1,becguard,databreaches}, they focus on the content in email messages, the infrastructure used to host phishing \textit{websites}, or involve attacks targeted primarily at \textit{consumers} (as opposed to enterprise organizations), leaving questions about the email delivery infrastructure unexplored.

To this end, we present the first large-scale study of the email delivery infrastructure used for enterprise phishing emails. This paper involves a collaboration between academic researchers and \cudaFullName, a large email security provider. We analyze the email delivery path information (from the email headers) across emails received by thousands of organizations. Our dataset consists of over 800,000 successfully delivered phishing emails and 4 billion non-phishing emails, across three different months (Jan 2020, Oct 2020, and Jan 2021). \enew{Importantly, the emails captured in our dataset have \textit{not been already blocked} by any pre-filtering services (e.g., Mimecast, Proofpoint) that organizations may employ, providing a representative view of real-world enterprise inboxes.}

Using this data, our paper investigates the following questions:
\begin{denseenum}
    \item What networks do enterprise phishing emails originate from?
    \item How does the email delivery infrastructure used in enterprise phishing attacks evolve over time?
    \item Can we use the delivery origin of an email to improve phishing detection? 
\end{denseenum}

Our analysis reveals several insights relevant to designing defenses. First, networks can be categorized by how much email sent from their servers is benign vs. phishing~(\S\ref{sec:ASsender}). On one end of the spectrum, the infrastructure of several prominent hosting companies (including Microsoft Azure and Amazon AWS) send a large volume of phishing emails, even though the vast majority of their outbound email is benign. These ``low-phishing-concentration" networks remain stable throughout the three months in our data.
On the other hand, a small number of IP addresses and networks send a large amount of phishing emails and only a small volume of benign emails. 
In contrast, this set of ``high-phishing-concentration'' networks is not stable over time~(\S\ref{sec:temporal}), suggesting that some subset of attackers actively switch to different networks to evade detection.


\enew{We apply the results of our empirical analysis in \cuda's production environment, where we monitor the rate and proportion of phishing emails coming from each network in an online fashion and demonstrate that we can use these dynamically updated features to detect previously undetected phishing attacks with a low false positive rate~(\cref{sec:detector}).}
Ultimately, the analysis and experiment results from our work suggest that we can better combat modern enterprise phishing by targeting the email delivery infrastructure they abuse through a diverse set of technical and policy-based defenses.

%% file: background.tex
\section{Background}
\label{sec:background}
Our paper studies the network characteristics of phishing emails received by enterprises.
We focus on emails sent from outside the recipient's organization, and make no assumptions about the kind of infrastructure or tools used by the attacker 
(\eg the emails might come from a compromised external account or reflect spoofing from an attacker-controlled server).
Below, we review the key networking-related information contained within email headers and provide an overview of related work.

\begin{figure}
    \centering
    \includegraphics[width=0.99\columnwidth]{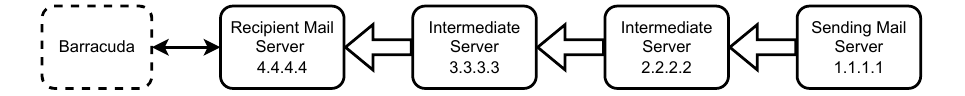}
    \caption{Example of email delivery with multiple mail servers. 
    \cuda collects and analyzes emails from the recipient's mail servers.} 
    \label{fig:emaildelivery}
    \vspace{-5ex}
\end{figure}

\subsection{Email Headers}
\label{sec:emailheaders}
In addition to their message content, emails contain a set of headers specified by the Simple Mail Transfer Protocol (SMTP).
When Alice sends an email, her mail client attaches \textit{envelope} headers
that specify the recipients' email addresses (\textit{RCPT TO}) and a return address for the sender (\textit{MAIL FROM}).
The email also contains human-readable headers that a recipient's email client will display as the email's sender and recipient (\textit{FROM} and \textit{TO} headers respectively). 

For each recipient, a copy of Alice's email will get routed through a series of mail servers (relays) until they arrive at the recipient's (Bob's) mail server.
Each relay will append a \textit{RECEIVED} Header to the email message that records information about the prior hop of the delivery path:
the IP address (and/or hostname) of the prior relay server and IP address (and/or hostname) of the current mail server receiving the email.
In the simplest case, Alice's mail server will transmit her email directly to Bob's mail server, without traversing any intermediate mail relays, resulting in a single \textit{RECEIVED} header.
However, in more complex scenarios, Alice's email might route through a series of mail servers, producing a set \textit{RECEIVED} headers. 
For example, Alice or Bob might use a large email provider that routes emails through a series of internal mail servers for scalability and load balancing purposes.
Figure~\ref{fig:emaildelivery} illustrates an example of email delivery that involves multiple mail servers, where the \textit{RECEIVED} headers might be: [Received, from 3.3.3.3 by 4.4.4.4], [Received, from 2.2.2.2 by 3.3.3.3], [Received, from 1.1.1.1 by 2.2.2.2].

The set of \textit{RECEIVED} headers describes the email \textit{relay path}, the underlying delivery infrastructure that an email traverses from sender to recipient.
We refer to the first (earliest) server in \textit{RECEIVED} header as the \textit{originating server} 
and its IP address as the \textit{originating IP address} (\eg 1.1.1.1 in Figure~\ref{fig:emaildelivery}).
Since \cuda collects and analyzes emails from the recipients' mail servers, our data contains the full set of headers after email delivery.

\subsection{Related Work}
\label{sec:related}
Below, we discuss several areas of research that are most relevant to this work.

\paragraph{Phishing Websites and URLs:}
Prior research proposes many methods to detect phishing websites and URLs~\cite{garera2007framework,prakash2010phishnet,zhang2007cantina,whittaker2010large,sheng2009empirical,tian2018needle,ma2009beyond} 
and analzyed the infrastructure used to host phishing websites~\cite{bitaab20:scampandamic,noroozian2019platforms}.
To detect malicious URLs, detectors can extract lexical features~\cite{garera2007framework,prakash2010phishnet}, or features based on a URL's domain~\cite{ma2009beyond}.
Prior work has also proposed machine learning detectors that use features relating to a website's content,
such as text from the rendered HTML DOM and images embedded in web pages~\cite{whittaker2010large,zhang2007cantina,tian2018needle,oest20:goldenhour}.
Unfortunately, phishing attacks have become increasingly sophisticated and employ various evasion strategies~\cite{phisheye,oest2020phishtime,zhang2021crawlphish,tian2018needle,databreaches}.
For example, recent studies indicate that attackers increasingly rely on compromised, legitimate websites 
or carefully craft their domain names to thwart URL-based detectors~\cite{desilva2021compromised,oest20:goldenhour}.

\paragraph{Phishing email detection:}
Prior work on detecting email attacks has predominantly used suspicious signals in the email's content~\cite{becguard,stringhini2015ain,gascon,duman2016emailprofiler,ho-2017,bergholz2008improved,lateral1}.
These approaches train machine learning models with features related to the email's text, URLs or attachments, and metadata such as the timing of the email and the consistency of its sender and recipient headers.
However, the continued success of phishing illustrates that existing defenses still have significant room to improve.

\paragraph{Email Authentication Protocols:} 
Several authentication protocols aim to combat spoofed emails. The Sender Policy Framework (SPF) allows a domain to add an allowlist to its DNS record that specifies the set of IP addresses allowed to send email on its behalf. In DKIM (DomainKeys Identified Email), domains add a public verification key to their DNS record and then append a private-key signature to emails they send. DMARC (Domain-based Message Authentication, Reporting, and Conformance) allows domains to specify a policy for how recipients should treat emails that fail to authenticate under either of the above protocols. While protocols can provide insight into the validity of a sender of an email, they cannot defend against other forms of deception, such as attacks that use compromised accounts or employ visually deceptive names. Furthermore, recent work shows it is very common for these authentication methods to break or be misconfigured~\cite{lazygatekeepersSPF,dmarc-enforcement}. 

\paragraph{Malicious email delivery infrastructure:}
Closer to our work, several studies have examined the email delivery infrastructure used in spam emails, a different and less deceptive email-based attack~\cite{ramachandran2006understanding,yardi2010detecting,qian2010network,john2009studying,hao2009detecting,lumezanu2012observing,xie2008spamming,sheng2009empirical}.
Ramachandran et al.~\cite{ramachandran2006understanding} found that a small number of autonomous systems (ASes) send the majority of spam and benign emails.
Building upon these results, prior work has found that using network-level information about an email's origin such as IP address and AS can improve spam detection both in machine learning models~\cite{hao2009detecting} and blocklists~\cite{qian2010network,FIRE}.
Furthermore, Fukushi et al.~\cite{fukushi2021comprehensive} found that thousands of IP addresses from popular cloud hosting services get blocked as a result of sending spam. They note that these IP addresses remain on blocklists for an average of 20--30 days, and due to the transient nature of cloud service IP addresses (e.g., machines shifting between users) this can create issues for benign users if such IP addresses remain on blocklists for too long.

\paragraph{Efficacy of modern blocklists:}
Recognizing the limitations of static blocklists and allowlists, some spam filters, such as SpamAssassin, assign a ``reputation'' score to received emails based on how many spam or benign emails were received in the past from their IP and email addresses~\cite{autowhitelist}. 
Although several modern blocklists (such as SURBL and Spamhause) receive frequent updates, they either do not include network-based features (in the case of SURBL) or focus primarily on spam (in the case of Spamhause), which is a different threat than the more targeted and deceptive emails used for enterprise phishing.
Furthermore, blocklists present several practical limitations: recent prior work~\cite{threat-intelligence} shows that the data sources of such blocklists are opaque and of questionable quality, and
the vendors providing these lists rarely explain their data collection and classification methodology.
Finally, blocklists fundamentally require someone to first report the attack, which not only introduces a delay \cite{sun2024victims}, but also means they cannot protect against unreported or previously unseen sources of attacks.\\

\noindent Although the prior work shows using network-level characteristics of an email's origin can help thwart spam,
it remains an open question whether modern phishing campaigns targeting enterprises rely on similarly positioned infrastructure and whether phishing exhibits other interesting network-level behaviors.

%% file: dataset.tex
\section{Data}
\label{sec:dataset}

Our dataset provides a unique view of enterprise phishing emails at scale, consisting of over 800,000 phishing emails and 4 billion clean emails received by enterprise users from external senders.
The data spans 3 one-month periods: January 2020, October 2020, and January 2021, and contains the SMTP headers of emails received by enterprises who use \cuda's services and have opted in to using their data for research purposes.
All the enterprises in our dataset use Microsoft Office 365 (O365) as their email provider.
Our dataset consists of emails that have been \textit{successfully} delivered to enterprise O365 accounts, so it does not contain emails that get blocked before delivery (e.g., by an organization's spam filter). Thus, our data reflects the typical phishing and clean emails that enterprise users actually encounter on a daily basis.

We analyze the effects of these email filtering services in~\S\ref{sec:no-filter} and show that organizations that deploy such filters do not encounter different phishing attacks as a result.
Then, in~\S\ref{sec:spoofing} we analyze the accuracy of the network origin of emails (\eg if attackers are forging \textit{RECEIVED} headers).
In~\S\ref{sec:authentication}, we compare the rates of email authentication protocols between clean and phishing emails. Finally, in~\S\ref{sec:limitations}, we discuss potential limitations with our dataset.
While we focus on analyzing the network characteristics of the phishing emails in our dataset, such a large-scale enterprise dataset could also be used to develop a phishing attack taxonomy or discover clusters of related attacks.

\subsection{Email Classification}
\label{sec:data collection}
We label emails as ``phishing'' or ``not-phishing'' via a set of \cuda's commercial detectors that detect phishing emails with embedded URLs. The detectors use text-based signals extracted from an email's body as well as features extracted from the embedded URL (\eg features from its DNS entry).
These classifiers do not incorporate features related to the network characteristics of an email and have an estimated precision of over 99\% for phishing emails that contain a link. 
The precision was calculated by a team of analysts at \cuda who manually analyze samples of emails labeled as attacks by the classifiers.
We also manually validated that a random sample of approximately 500 emails (from distinct campaigns) labeled as phishing within our dataset were indeed phishing emails by inspecting the subject line, email body, and sender information.
Given the scale of our dataset (which prohibits manual labeling) and low false positive rate of these classifiers, we treat \enew{\cuda's classifiers'} labels as ground truth.
Because our work focuses on the characteristics of enterprise phishing emails, we refer to non-phishing emails as benign or ``clean'', although a small subset of these emails might correspond to phishing or other forms of abusive emails (such as extortion and scams).
As we show in \S\ref{sec:detector}, despite the existence of false negatives in our dataset, we identify a set of useful features related to the network origins of emails that helps uncover new, previously unclassified phishing emails.

\begin{table}
\centering
\begin{tabular}{|l|r|r|r|}
\hline
\textbf{Statistic}                          & \textbf{Jan 2020} & \textbf{Oct 2020} & \textbf{Jan 2021} \\ \hline
Phishing emails                   & 218,079            & 307,279            & 282,689            \\  \hline
Phishing campaigns                       & 67,176             & 80,703             & 73,925             \\  \hline
Unique IP addresses (phishing)                      & 16,027             & 20,596             & 12,743             \\  \hline
Unique ASes (phishing)                     & 2,808              & 3,317              & 2,075              \\  \hline
Clean emails                      & $\sim$2B   & $\sim$1B   & $\sim$1.3B \\ \hline
\end{tabular}
\caption{Aggregate statistics of dataset.}
\vspace{-7ex}
\label{table:gen-data-stats}
\end{table}

\subsection{General Statistics}
\label{sec:dataset-desc}
Table~\ref{table:gen-data-stats} summarizes the size of our data set.
Each month in our dataset contains 200,000 - 300,000 phishing emails and over 1 billion clean emails.
Phishing emails originate from over 12,000 distinct IP addresses across each month of data (row 3).
For our analysis, we also mapped IP addresses to the autonomous system (AS) they belong to using the Cymru IP address to ASN mapping API~\cite{cymruAS}.

\begin{figure}[t]
    \centering
    \includegraphics[width=0.6\columnwidth]{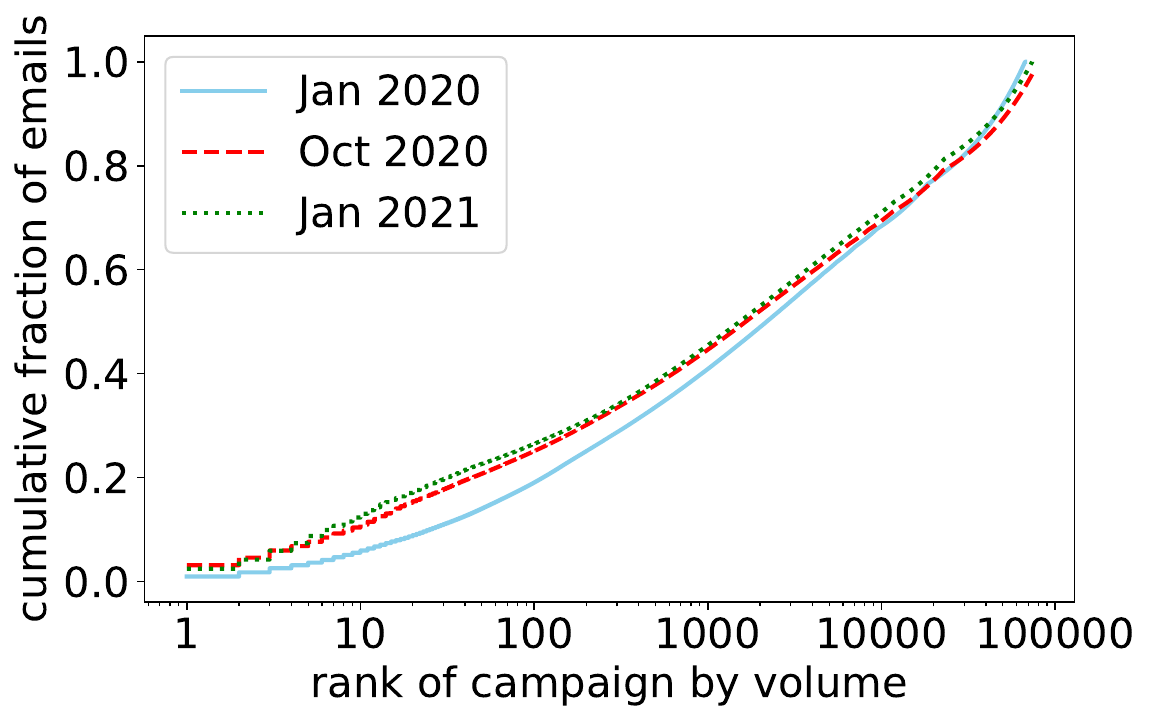}
    \caption{Cumulative fraction of phishing emails per campaign.}
    \vspace{-3ex}
    \label{fig:num-email-campaign}
\end{figure}

\paragraph{Campaign analysis:}
We computed the number of distinct campaigns in our dataset to understand whether the phishing emails in our data reflect a range of different attacks.
Following prior work~\cite{lateral1}, we define a phishing campaign as a set of phishing emails that contain the same \textit{FROM} address and normalized subject line (lower-cased with spaces and symbols removed). 
Figure~\ref{fig:num-email-campaign} shows the cumulative fraction of phishing emails, as a function of each campaign, ranked by phishing volume.
Although the top-sending campaigns contribute significant volume, (\eg the top-1000 campaigns account for over 40\% of the email volume in all three periods), we find that the distribution is heavy-tailed and conclude that no single campaign is skewing our dataset. 
While our work mainly focuses on high-volume phishers, future work could explore smaller campaigns to reveal patterns and threats in more targeted attacks.

\paragraph{Data Cleaning:} 
We first removed all \textit{RECEIVED} headers whose IP address fell within a reserved or private address range. Then, we extracted the origin IP address (as defined in ~\S\ref{sec:emailheaders}) from the remaining set of headers (\ie from the first header without a private or reserved IP address).

\subsection{Effect of Email Filtering Services}
\label{sec:no-filter}
\enew{Some organizations employ additional email filters (\eg spam filters or blocklists) that prevent some emails from reaching their users' inboxes (and thus are not captured by our dataset). In this section, we measure the prevalence and effect of these ``pre-filtering'' services on our dataset.}
\enew{We find that 75\% of organizations use some pre-filtering services (e.g., a spam filter from Mimecast or Barracuda). To determine whether an organization uses a pre-filtering service, we analyzed their MX record: If their MX record points directly to Microsoft Office 365's IP addresses, then the organization likely does not use an external filtering service, and all email will be delivered to users' O365 accounts.\footnote{Email classified as spam by Microsoft will still appear in a user's account, but within a spam folder (which our dataset includes).}} 
To understand the impact of pre-filtering on our dataset,
we examined whether organizations that do not pre-filter their emails encounter uniquely different phishing attacks as a result (as opposed to natural variations in phishing attacks received by different enterprises).

Using the data from Jan 2020, we compared the phishing attacks received only by organizations without pre-filtering (55,515 emails) to attacks at organizations that do apply pre-filters (162,564 emails).
Across organizations with no filtering, 41,474 emails originated from an IP address that also sent phishing emails to organizations with filters, while 14,041 emails originated from IP addresses that were only seen in phishing emails at organizations without filtering.
We checked the inverse relationship by comparing the phishing attacks received only by organizations without pre-filtering to attacks received only by an equally-sized random sample of organizations that do apply pre-filters (54,871 emails).
Interestingly, we found the number of phishing emails received only by organizations with pre-filtering was similar at 14,823. These numbers did not significantly change when comparing at the AS granularity, or when we grouped the emails by phishing campaign.

Thus, we find that the majority of phishing emails in our dataset originate from IP addresses observed in the inboxes of both organizations that use pre-filtering services and organizations that do not. While both sets of organizations receive some phishing emails from distinct sets of origin IP addresses, the volume of phishing emails from the distinct origin IP addresses is comparable.
These results suggest that 1) a significant number of phishing attacks still bypass traditional email filters; and 2) our dataset provides a representative view of phishing emails received by real-world enterprises despite the use of email pre-filtering.

\subsection{Accuracy of Network Origin}
\label{sec:spoofing}
\enew{In this section, we analyze the fidelity of the \textit{RECEIVED} headers in phishing emails. In particular, an attacker could obfuscate their origins by adding additional arbitrary \textit{RECEIVED} headers to the start of an email's delivery path. However, the recipient's mail server generates the last set of \textit{RECEIVED} headers, including the IP address of the server that delivered the email directly to the recipient's server. Thus, we can reliably determine the email relay immediately preceding the recipient's mail server.}

We extract the \textit{RECEIVED} header that contains reliable sender information as follows: We first find the header that contains the IP address of the recipient's mail server based on the recipient's DNS MX record.\footnote{To avoid additional evasion, if multiple headers claim to involve recipient server, our analysis only uses the last of these to ensure it reflects the true recipient server's information. We also ensured that we used the historical DNS MX record from the corresponding time period the email was delivered.}
From this header, we then know that the previous \textit{RECEIVED} header was generated by the recipient's mail server. As a result, this header reliably reports the IP address of the server from which it received the email.

Based on a series of tests comparing reliable email relay information and potentially forged information, we find that at least 90.6\% of phishing emails in our dataset do \textit{not} spoof the origin IP address in their \textit{RECEIVED} headers (at least not in ways that impact our analysis). For additional information on the tests, see Appendix \ref{appendix:spoofing-validation}, which suggests that for the vast majority of phishing emails, we do not see strong evidence of email delivery path forgery.

\subsection{Email Authentication}
\label{sec:authentication}
Given the low implementation and high misconfiguration rates of email authentication protocols found by prior work (such as \cite{lazygatekeepersSPF}) we investigated whether email authentication rates differed between clean and phishing emails within our dataset. Table~\ref{table:spf-dkim-dmarc-results} compares the proportion of phishing and clean emails that pass various authentication checks in our data in January 2020; 
these statistics remain consistent in October 2020 and January 2021. 
Although clean emails pass at a higher rate than phishing emails, less than half of clean emails pass DMARC validation, and 10.4\% of phishing emails pass DMARC checks.
Thus, while we find that email authentication rates differ slightly between phishing and clean emails, they fail to provide a reliable signal for detecting phishing. We confirm this finding in \S\ref{sec:detector}, where we find that email authentication protocols are the least important features when distinguishing between clean and phishing emails.

\begin{table}[t]
\centering
\begin{tabular}{lrr}
\multicolumn{1}{l}{\textbf{Protocol: Pass Rate}} & \multicolumn{1}{l}{\textbf{Clean}} & \multicolumn{1}{l}{\textbf{Phishing}} \\ \hline
\multicolumn{1}{l}{\multirow{1}{*}{\textbf{SPF or DKIM}}}   & 39.6\% & 23.3\%                  \\
\multicolumn{1}{l}{\multirow{1}{*}{\textbf{SPF and DKIM}}}  & 10.0\% & 10.1\%                  \\ 
\multicolumn{1}{l}{\multirow{1}{*}{\textbf{DMARC}}} & 30.0\%      & 10.4\%  \\
\end{tabular}
\caption{Proportion of emails that pass authentication in Jan '20.
}
\label{table:spf-dkim-dmarc-results}
\vspace{-5ex}
\end{table}

\subsection{Limitations}
\label{sec:limitations}
Although we empirically determine that the usage of pre-filtering does not significantly bias our dataset~(\S\ref{sec:no-filter}), we acknowledge that using MX records to infer the usage of pre-filtering may be imperfect. Specifically, some organizations may use multiple filtering layers not detectable through solely inspecting their MX records (e.g., if they route email through security apps within O365).
Other biases may arise from the fact that our dataset consists predominantly of organizations based in the U.S. and Europe (so it may not be reflective of phishing emails targeting other parts of the globe). 
Our dataset only captures data up to 2021 and may not reflect any changes in the landscape of email phishing since then.
Additionally, although our dataset consists of billions of emails, it contains only a small number of phishing emails that come from major email service providers, such as Gmail or Outlook\footnote{In total, $\sim$8 million emails originate from Google's address space, and fewer than 50 phishing emails originate from published Gmail or Outlook IP address ranges. }. 
A potential explanation is that major email providers have tight security controls on the outbound emails from accounts they manage, which limits the type (e.g., emails with a spoofed \textsc{From} address) and volume (i.e., rate limits on outbound emails~\cite{gmail-ratelimit}) of emails that can be sent by attackers that use accounts on these services.

\subsection{Ethics}
\label{sec:ethics}
The characterization was conducted on an existing dataset of email headers from organizations who are active customers of \cudaFullName. Per Barracuda’s policies, all fetched emails were stored encrypted. 
Only the researchers working on this project and authorized employees at \cuda were allowed to access the data, under a data-sharing agreement with the researchers’ institution and via standard, strict access control policies. 
The academic researchers analyzing this data submitted this study and received approval from their institution's IRB. See Appendix \ref{appendix:ethics} for further discussion on our ethical considerations.

%% file: results.tex
\section{Delivery Infrastructure Characterization}
\label{sec:results}

In this section, we investigate the ASes and IP addresses of the email delivery infrastructure the emails in our dataset. First, we examine the distribution of phishing emails sent per IP and AS (\S4.1). We find that a small set of IP addresses accounts for a substantial portion of phishing emails. We then characterize AS behaviors from two perspectives: the fraction of emails sent from an AS that are phishing (\S4.2) and the stability of the set of ASes utilized for phishing attacks (\S4.3). We conclude with an analysis of the geographical distribution of the email delivery infrastructure (\S4.4).

\begin{figure}[t]
\begin{minipage}[t]{0.5\linewidth}
    \centering
    \includegraphics[width=1\textwidth]{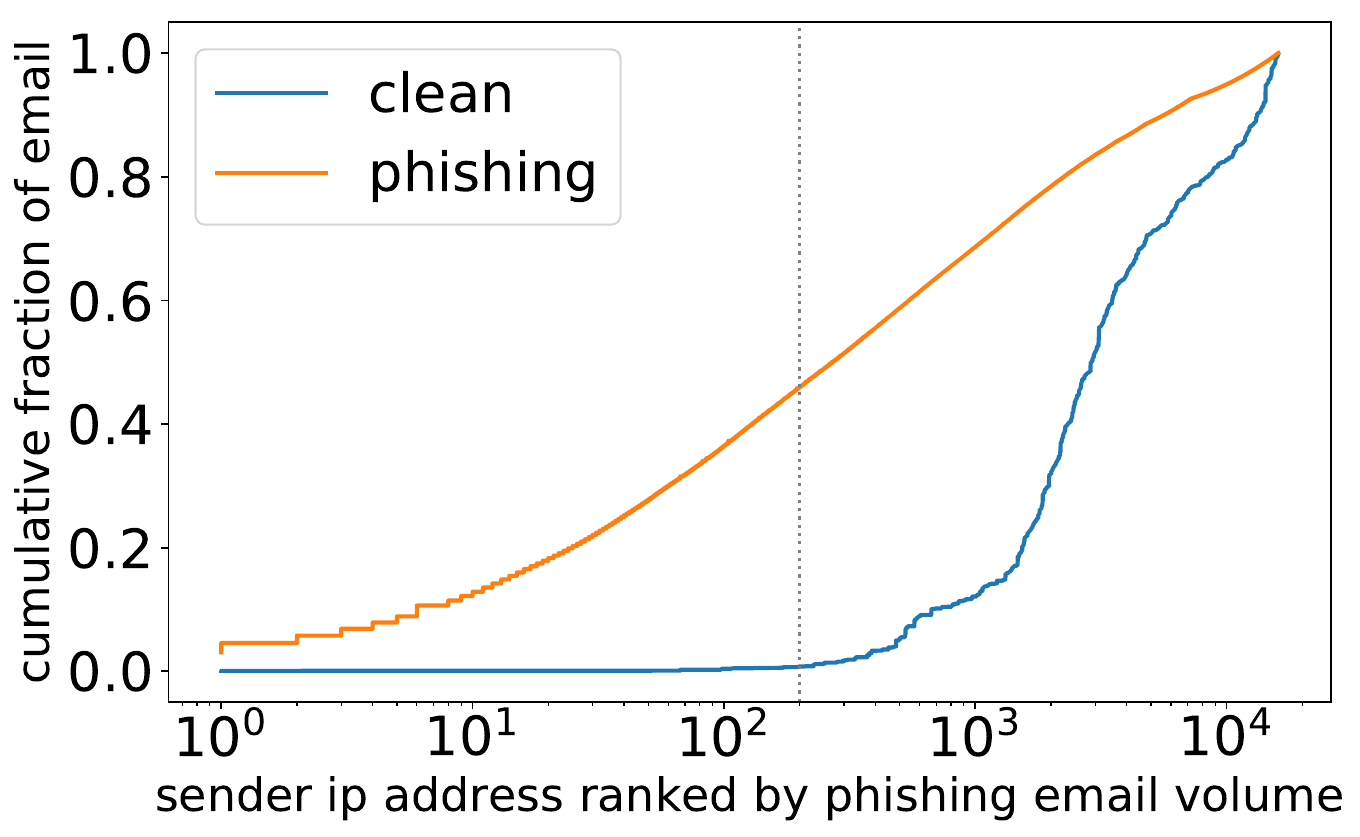}
    \caption{Cumulative fraction of phishing emails by originating IP address. The 200 highest-volume sending IP addresses account for around 50\% of phishing emails.}
    \label{fig:cumulativePhisingPerIP}
\end{minipage}
\hspace{0.5cm}
\begin{minipage}[t]{0.5\linewidth} 
    \centering
    \includegraphics[width=1\textwidth]{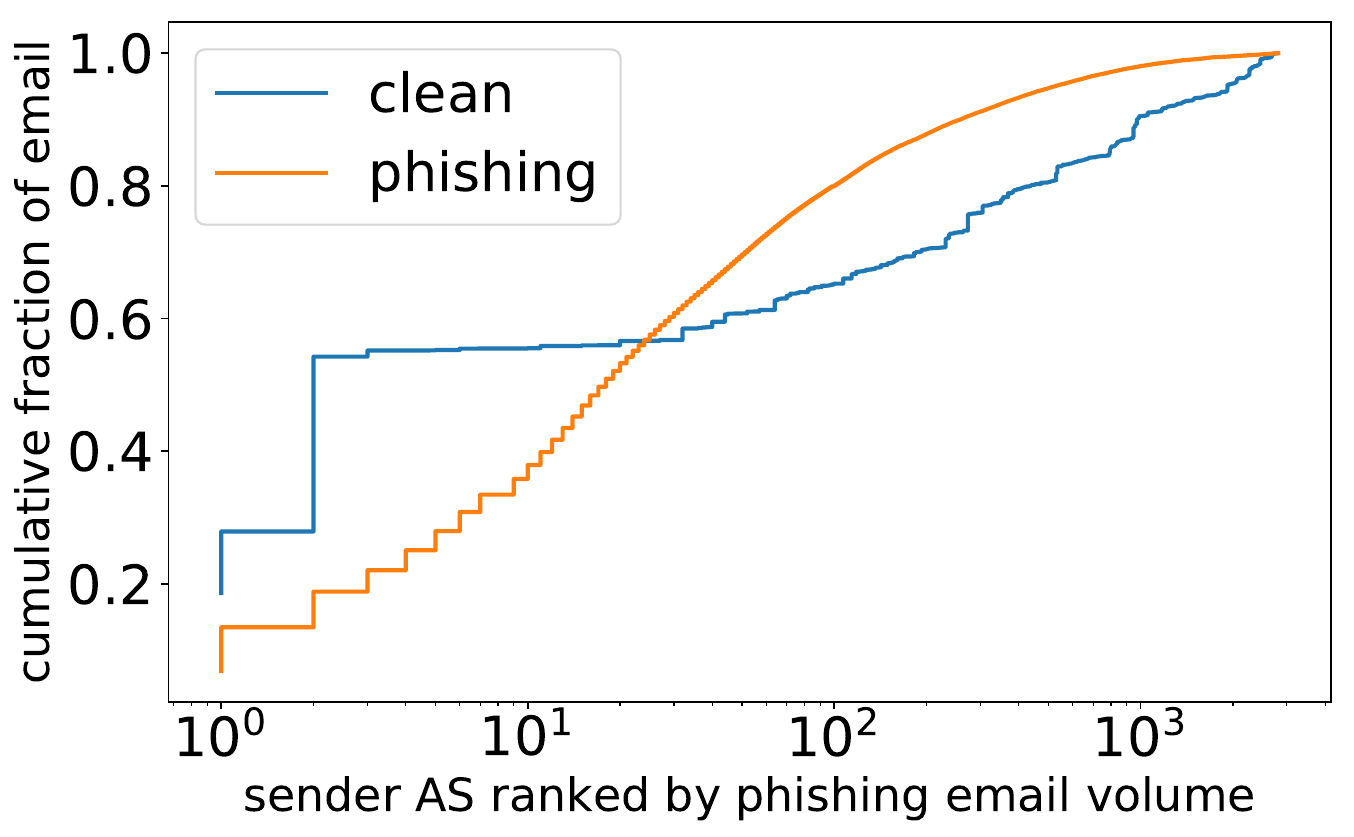}
    \caption{Cumulative fraction of phishing emails sent across the ASes of an email's originating IP address.}
    \label{fig:cumulativePhishingPerAS}
\end{minipage}
\vspace{-2ex}
\end{figure}

\subsection{Network Origin Distribution of Phishing Emails}
\label{sec:ASsender}
In this section, we find a significant volumes of phishing comes from a handful of ASes and IP addresses that do not send large amounts of clean emails, suggesting that network origin reputation features, extracted from the email delivery path, can be a useful signal for detecting phishing emails.

\paragraph{Sender IP address distribution:}
\Cref{fig:cumulativePhisingPerIP} shows the cumulative fraction of clean and phishing emails sent per source IP address, sorted by the number of phishing emails each sent. 
Although our dataset includes phishing emails successfully delivered from over 10,000 unique IP addresses, the distribution is very skewed, with around 50\% of the phishing emails in our data coming from the top 200 IP addresses alone. These IP addresses account for less than 1\% of the clean emails. This result is in contrast to prior work on spam~\cite{ramachandran2006understanding}, which found that, while a small set of IP addresses accounted for a significant amount of spam, the majority of clean emails also originated from this same address set.

\paragraph{AS distribution:} 
We illustrate the distribution of phishing emails per AS in Figure~\ref{fig:cumulativePhishingPerAS} (where the $x$-axis is sorted by the volume of phishing emails that originated from each AS). We find that approximately 80\% of the phishing emails in our dataset come from just 100 ASes, which represent fewer than 1\% of all ASes. This same set of 100 ASes accounts for nearly 70\% of the clean emails in our dataset. Nearly 60\% of clean emails in this set originate from the top three ASes (2 belonging to Amazon and 1 belonging to Microsoft). Excluding Amazon and Microsoft, the top-100 ASes still account for the majority of phishing emails in our data, but less than 10\% of the clean emails.

\paragraph{Takeaways:}
\enew{These findings suggest that network-based reputation features can help detect previously undetected phishing attacks from some networks (e.g., not Amazon or Microsoft). In subsequent sections, we explore additional aspects about the networks sending phishing emails that can bolster the performance of such features.}

\begin{figure}[t]
    \centering
    \includegraphics[width=0.7\columnwidth]{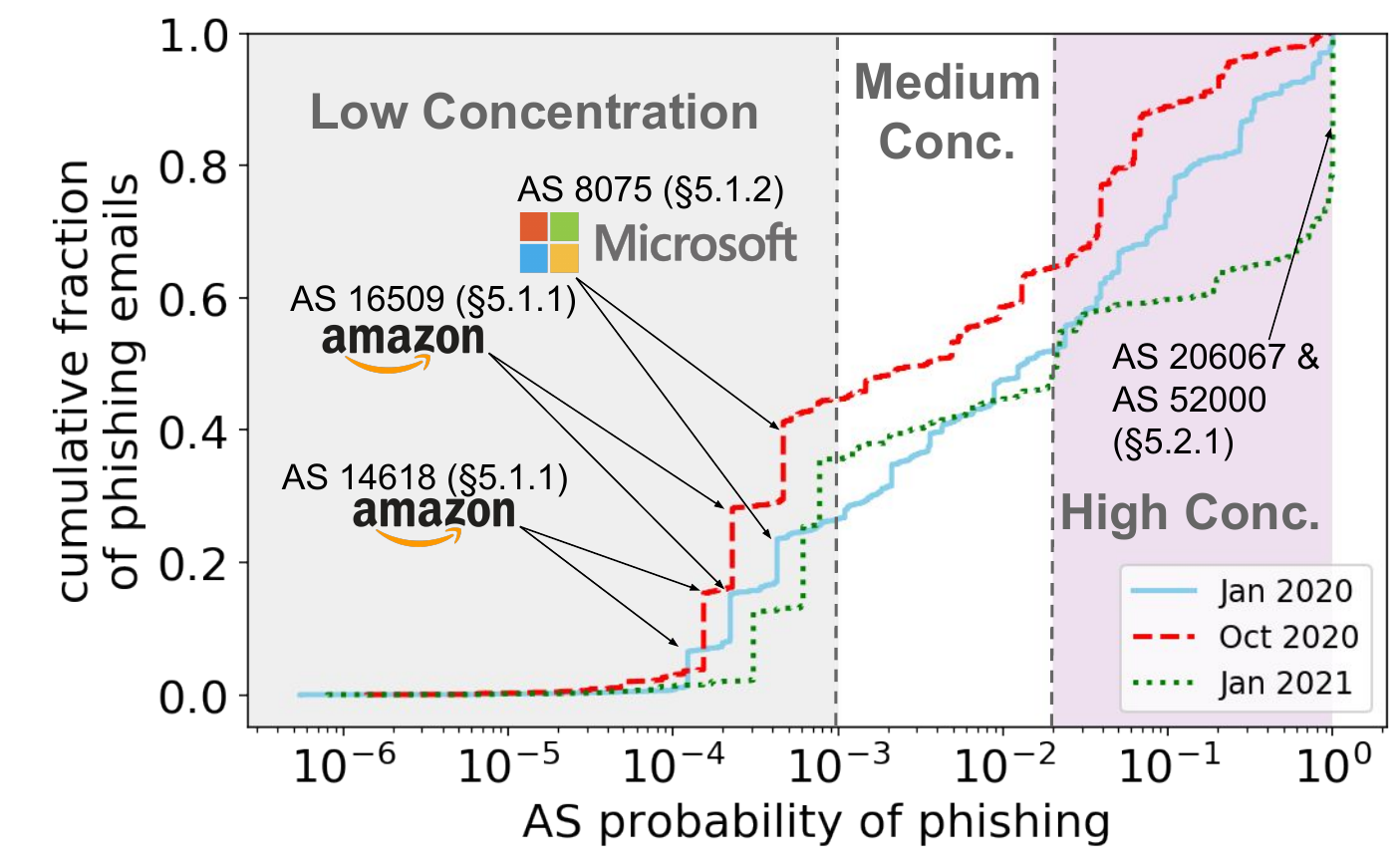}
    \caption{Cumulative fraction of phishing emails as a function of phishing probability of an AS, including AS phishing concentration categories. In January 2021, AS 206067 and AS 52000 had probabilities of phishing of 99.8\% \enew{(with 6550 delivered phishing emails)} and 100\% \enew{(with 1673 delivered phishing emails)}, respectively.} 
    \label{fig:categoriesCDF}
    \vspace{-3ex}
\end{figure}

\subsection{Phishing Concentration Across Networks}
\label{sec:infrastructure}
\enew{To characterize variations in behavior of the networks sending phishing emails, we divided the ASes in our dataset into three categories, low, medium, and high phishing concentration, using the following metrics: (1) the volume of delivered phishing email originating from each network; and (2) the AS's \textit{probability of phishing}, defined as the fraction of delivered phishing emails originating from a network divided by the total number of emails (clean and phishing) from that network. The heuristics for each category are as follows:} 
\begin{itemize}
    \item \textbf{Low phishing concentration network:} an AS where less than 0.1\% of its sent emails are phishing.
    \item \textbf{Medium phishing concentration network:} an AS where more than or equal to 0.1\% and less than 2\% of its sent emails are phishing.
    \item \textbf{High phishing concentration network:} an AS with more than or equal to 2\% of its sent emails labeled as phishing and has sent more than 150 phishing emails in one month. This 150-email threshold removes any ASes that sent very few total emails, which could lead to skewed probability values.\footnote{ASes that sent less than 150 emails collectively account for 5.5\% of all phishing emails in January 2020, 2.7\% in October 2020, and 5\% in January 2021.}
\end{itemize}

We chose to focus on the AS-granularity to reveal coarser-grain patterns (e.g., large volumes of phishing emails from reputable ASes). We further explore more fine-grained patterns (i.e., at the IP address level) in our case studies~(\S\ref{sec:case:low_risk}).

Table~\ref{tab:num-networks-category} summarizes the proportion and size of each phishing concentration category, and Figure~\ref{fig:categoriesCDF} shows the distributions of phishing emails across networks' probability of phishing. This distribution is similar for each of the three time periods, with a few low phishing concentration (but high phishing volume) networks contributing to nearly one-third of all phishing email.

\begin{table}[t]
\centering
\resizebox{\textwidth}{!}{ 
\begin{tabular}{|l|cc|cc|cc|c|}
\hline
\multicolumn{1}{|c|}{\multirow{2}{*}{\textbf{Category}}} & \multicolumn{2}{c|}{\textbf{Jan 2020}}                                                                                                      & \multicolumn{2}{c|}{\textbf{Oct 2020}}                                                                                                      & \multicolumn{2}{c|}{\textbf{Jan 2021}}                                                                                                      & \textbf{Overall}                                                 \\ \cline{2-8} 
\multicolumn{1}{|c|}{}                                   & \multicolumn{1}{c|}{\begin{tabular}[c]{@{}c@{}}\# of\\ ASes\end{tabular}} & \begin{tabular}[c]{@{}c@{}}\% Phishing\\ Emails Sent\end{tabular} & \multicolumn{1}{c|}{\begin{tabular}[c]{@{}c@{}}\# of\\ ASes\end{tabular}} & \begin{tabular}[c]{@{}c@{}}\% Phishing\\ Emails Sent\end{tabular} & \multicolumn{1}{c|}{\begin{tabular}[c]{@{}c@{}}\# of\\ ASes\end{tabular}} & \begin{tabular}[c]{@{}c@{}}\% Phishing\\ Emails Sent\end{tabular} & \begin{tabular}[c]{@{}c@{}}\% Phishing\\ Emails Sent\end{tabular} \\ \hline
Low Concentration                                        & \multicolumn{1}{c|}{608}                                                 & 26.58                                                            & \multicolumn{1}{c|}{537}                                                 & 44.64                                                            & \multicolumn{1}{c|}{311}                                                 & 35.66                                                            & 35.63                                                            \\ \hline
Medium Concentration                                     & \multicolumn{1}{c|}{1031}                                                & 25.39                                                            & \multicolumn{1}{c|}{502}                                                 & 19.94                                                            & \multicolumn{1}{c|}{397}                                                 & 13.82                                                            & 19.73                                                            \\ \hline
High Concentration                                       & \multicolumn{1}{c|}{62}                                                  & 42.5                                                             & \multicolumn{1}{c|}{44}                                                  & 32.76                                                            & \multicolumn{1}{c|}{61}                                                  & 43.72                                                            & 39.66                                                            \\ \hline
\end{tabular}
}
\caption{Number of ASes for which a low ($<$0.1\%), medium (0.1\%-2\%), or high ($\ge$2\%) proportion of their emails are phishing, and percent of phishing emails originating from each category.}
\vspace{-3ex}
\label{tab:num-networks-category}
\end{table}

\paragraph{Low phishing concentration networks:}
Three networks belonging to Amazon (AS 16509 and AS 14618) and Microsoft (AS 8075) are responsible for over 85\% of all phishing emails from networks in this category, and 31\% of all phishing email in our dataset. 
Despite the large amount of phishing emails originating from their address space, these ASes are classified as low phishing concentration as they are also the source of tens of millions of legitimate emails (Table~\ref{tab:large-volume-low-risk}), accounting for 60\% of all legitimate email (as seen in the three large jumps in Figure~\ref{fig:cumulativePhishingPerAS}).

\begin{table}[t]
\centering
\footnotesize
\begin{tabular}{|l|r|r|r|}
\hline
\textbf{ASN} &
  \multicolumn{1}{l|}{\textbf{Date}} &
  \multicolumn{1}{l|}{\textbf{\# Phishing}} &
  \multicolumn{1}{l|}{\textbf{\# Clean}} \\ \hline
\multirow{3}{*}{14618 (Amazon)}   & Jan 2020 & 6,144  & 49,655,611 \\ \cline{2-4} 
                                     & Oct 2020 & 10,574 & 68,629,830 \\ \cline{2-4} 
                                     & Jan 2021 & 12,518 & 40,778,516 \\ \hline
\multirow{3}{*}{16509 (Amazon)}   & Jan 2020 & 7,845  & 35,086,658 \\ \cline{2-4} 
                                     & Oct 2020 & 10,902 & 46,979,360 \\ \cline{2-4} 
                                     & Jan 2021 & 14,465 & 23,541,485 \\ \hline
\multirow{3}{*}{8075 (Microsoft)} & Jan 2020 & 7,507  & 17,481,559 \\ \cline{2-4} 
                                     & Oct 2020 & 10,727 & 22,886,922 \\ \cline{2-4} 
                                     & Jan 2021 & 11,839 & 15,272,305 \\ \hline
\end{tabular}
\caption{Email volume over time of the top 3 low phishing concentration networks.}
\vspace{-5ex}
\label{tab:large-volume-low-risk}
\end{table}

\paragraph{Medium phishing concentration networks:}
This category accounts for the lowest share of phishing emails (19.7\%) and includes some smaller Internet hosting companies. These networks are largely used to send benign email, but still originate an higher than average proportion of phishing.

\paragraph{High phishing concentration networks:}
ASes in this category typically send a relatively small number of emails overall but a high proportion of phishing emails. \enew{This category contains the networks responsible for the largest amount of phishing email in our dataset: over 39.7\%.}
We find that many high phishing concentration networks are legitimate hosting companies. To classify the networks, we visited the companies' webpages and validated our classifications by comparing them to CAIDA's inferences~\cite{caida-as-classification} and to ASes' self-classifications on PeeringDB~\cite{peeringdb-dot-com} (when available). In most cases, our classification of an AS as a hosting provider was validated by PeeringDB and/or CAIDA classifying it as a \textsc{Content} provider. Most cases of disagreement were due to the AS offering \textit{both} hosting and consumer Internet services.

We found that on average, 67.6\% of networks we classified as high-concentration provide hosting services.
Aside from Internet hosting companies, the vast majority of the remaining ASes are Internet service providers, with only 1-2\% belonging to other AS types, such as \textsc{Enterprise} or \textsc{Educational/Research}.
We suspect many of these networks are unwitting hosts of attackers, who quickly send a large amount of phishing email before moving on to new network infrastructure (e.g., after getting caught), although some could be bullet-proof hosting providers that use resources from legitimate upstream providers \cite{noroozian2019platforms}. We examine some of these "transient" malicious networks in more detail in our case studies (\S\ref{sec:caseStudies}).

\paragraph{Takeaways:}
Using network-based reputation features may be more useful for thwarting phishing emails originating from high phishing concentration networks. On the other hand, since low phishing concentration networks originate a large amount of legitimate email traffic, such reputation features will be unlikely to accurately capture all phishing activity (e.g., Amazon/Microsoft IP addresses are unlikely to show up on fraud-activity-based blocklists).

\subsection{How Does Phishing Delivery Infrastructure Change Over Time?}
\label{sec:temporal}
\enew{In this section, we characterize the stability the email delivery infrastructure used to send enterprise phishing emails. In particular, we examine the stability of high phishing concentration networks and the top-100 ASes by phishing volume rank over a period of one year.}

\begin{figure}[t]
    \centering
    \includegraphics[width=0.7\columnwidth]{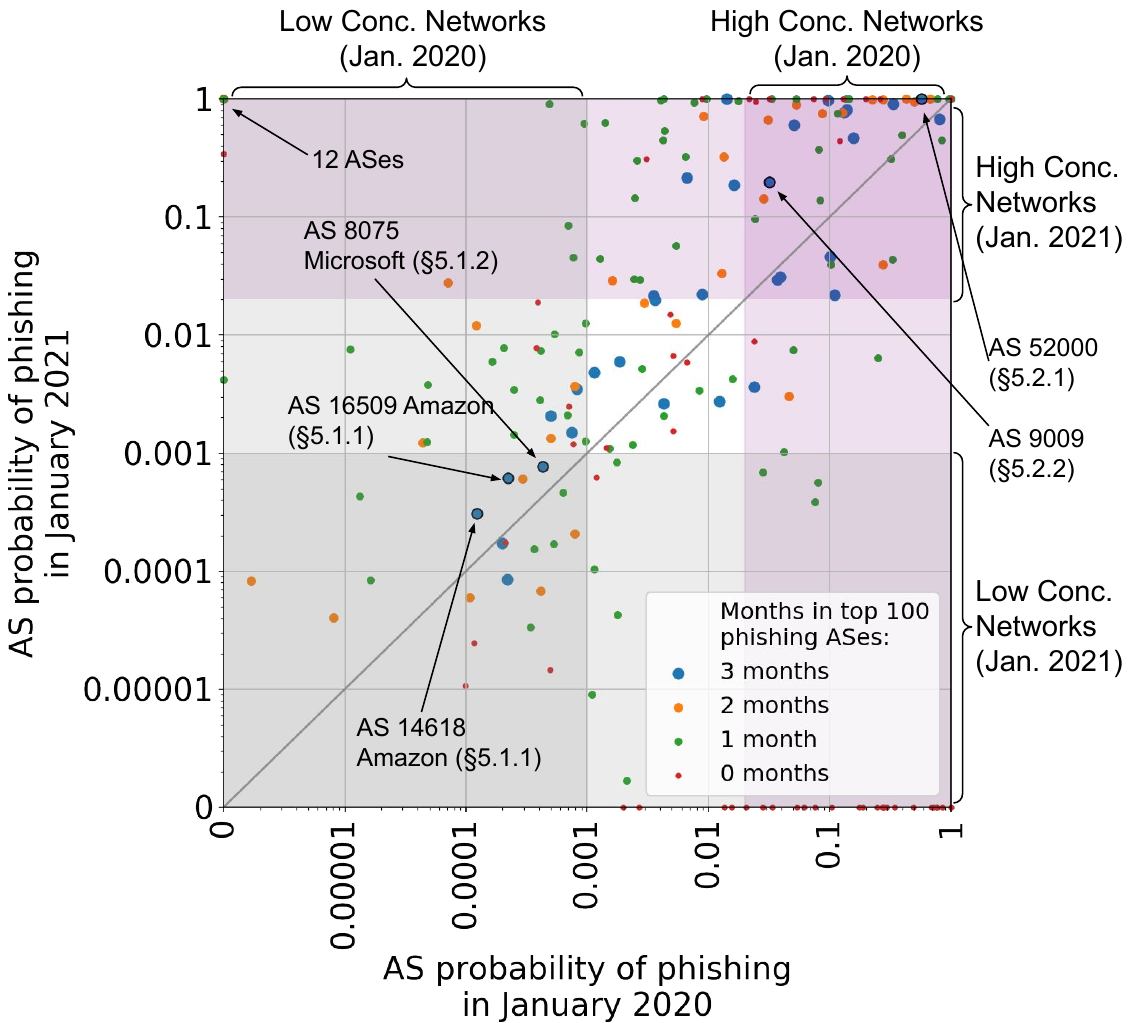}
    \caption{AS probability of phishing in Jan '20 and '21. Each point is AS that sent 100 or more phishing emails in \textit{either} Jan '20 or Jan '21. Purple regions correspond to high phishing concentration ASes.} 
    \vspace{-3ex}
    \label{fig:probOverTime}
\end{figure}

\paragraph{Year-to-year comparison of phishing probability:}
\enew{Comparing the fraction of an AS's emails to \cuda customers that are phishing in January 2020 ($X$ axis) to the fraction that are phishing in January 2021 ($Y$ axis) (for ASes that sent at least 100 phishing emails), we find that most ASes fall fairly close to the line $X=Y$, meaning they maintain a consistent probability of phishing across the two time periods (\Cref{fig:probOverTime}).} The ASes in the top-right corner of \Cref{fig:probOverTime} have a consistently high probability of phishing, although many transition from being very likely to send phishing emails in 2020  (compared to the global fraction of email that is phishing) to sending exclusively phishing emails in 2021. We find that all ASes with a probability of phishing of 0 in either time frame did not send emails during that month, meaning they were actively used by attackers in one year and did not send any emails during the other month to organizations in our dataset. 

\begin{table}[t]
\centering
\resizebox{0.65\textwidth}{!}{ 
\begin{tabular}{rl|r|r|r|}
\cline{3-5}
\multicolumn{1}{l}{}         &                 & \multicolumn{3}{c|}{\textbf{Phishing Email Volume}} \\ \hline
\multicolumn{1}{|l|}{\textbf{ASN}} &
  \textbf{Owner} &
  \multicolumn{1}{l|}{\textbf{Jan 2020}} &
  \multicolumn{1}{l|}{\textbf{Oct 2020}} &
  \multicolumn{1}{l|}{\textbf{Jan 2021}} \\ \hline
\multicolumn{1}{|r|}{4808}   & CHINA169-BJ     & 2143            & 2405            & 2163            \\ \hline
\multicolumn{1}{|r|}{9009}   & \textbf{M247 Ltd}        & 955             & 3207            & 2135            \\ \hline
\multicolumn{1}{|r|}{31863}  & \textbf{Centrilogic}     & 557             & 605             & 1141            \\ \hline
\multicolumn{1}{|r|}{40676}  & \textbf{Psychz Networks} & 1974            & 210             & 618             \\ \hline
\multicolumn{1}{|r|}{54290}  & \textbf{Hostwinds}       & 2422            & 1340            & 401             \\ \hline
\multicolumn{1}{|r|}{60068}  & \textbf{CDN77}           & 414             & 459             & 373             \\ \hline
\multicolumn{1}{|r|}{64236}  & \textbf{Unreal Servers}  & 2014            & 353             & 372             \\ \hline
\multicolumn{1}{|r|}{135905} & VNPT-AS-VN      & 581             & 274             & 355             \\ \hline
\multicolumn{1}{|r|}{197226} & \textbf{SPRINT S.A.}     & 636             & 332             & 288             \\ \hline
\end{tabular}
}
\caption{The 9 persistent high phishing concentration networks and their phishing email volume over our three datasets. Networks that provide hosting services are bolded.}
\vspace{-5ex}
\label{table:persistent-high-risk-networks}
\end{table}

\begin{figure}[t]
    \centering
     \begin{subfigure}{0.5\columnwidth}
    \includegraphics[width=0.70\textwidth]{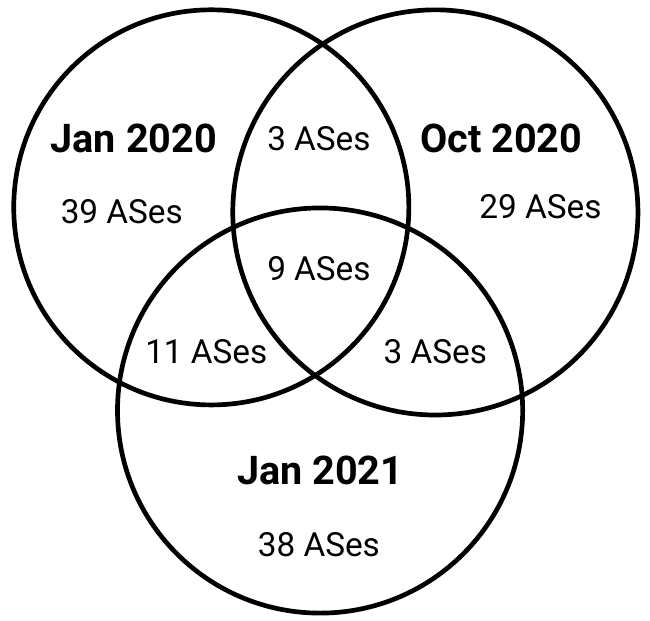}
    \subcaption{High phishing concentration ASes.} 
    \label{fig:venn-high-risk}
\end{subfigure}~
\begin{subfigure}{0.5\columnwidth}
    \includegraphics[width=0.70\textwidth]{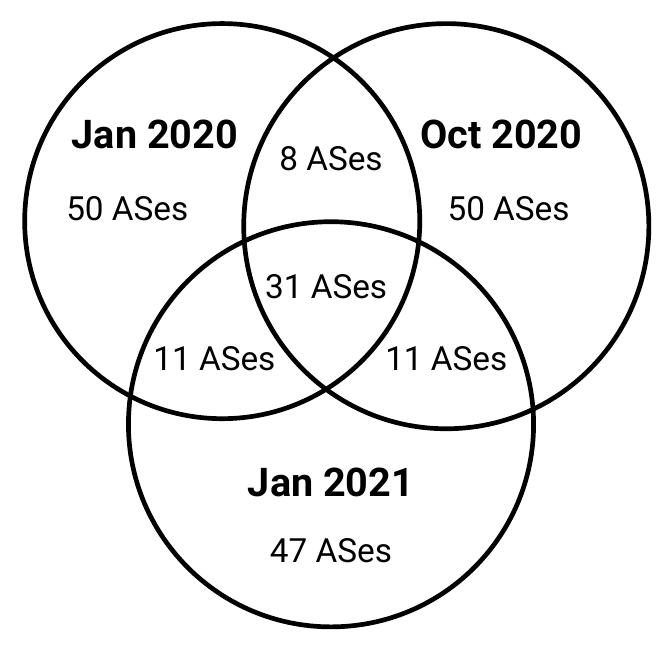}
    \subcaption{Top-100 ASes by phishing volume.}
    \label{fig:venn-top-100}
    \end{subfigure}
    \caption{Persistence of networks over time.}
    \vspace{-2.5ex}
\end{figure}

\paragraph{Stability of high phishing concentration networks:}
\enew{Of the 132 ASes we classify as high phishing concentration during at least one of the months in our dataset, the majority are only high-concentration for one month (Figure~\ref{fig:venn-high-risk}). Only 26 ASes (19.6\%) are classified as high phishing concentration in at least two out of three months of our data, and 9 ASes (6.8\%) are high-concentration across all three months. When combined, over 25,000 phishing emails originate from the 9 persistent high phishing concentration networks across the three months in our data (Table~\ref{table:persistent-high-risk-networks}).}

\paragraph{Stability of top-100 ASes by phishing email count:} 
\enew{Out of the top-100 networks by phishing volume, many consistently rank among the top-100 over time: 31 remain across all three months in our data, and 61 ASes appear in the top-100 phishing networks across at least two months (Figure~\ref{fig:venn-top-100}). Overall, the top-100 ASes by phishing volume account for 83\% of the total volume of delivered phishing emails. The 31 ASes that are consistently in the top-100 account for 44\%-51\% of the total phishing email, and the 61 ASes account for 49\%-65\%. Of the 31 persistent networks, 26 provide hosting services, while the remaining 5 ASes are Internet service providers.}

\paragraph{Takeaways:}
While some networks are consistently reused to send phishing emails, a larger proportion of phishing email delivery infrastructure is comparatively transient. These nuances suggest that incorporating temporal aspects of network behavior may help improve phishing detection while maintaining a low false positive rate. In Section~\ref{sec:caseStudies}, we investigate these nuances in more detail and discuss defensive ideas that leverage our insights. We empirically confirm this idea in a real-world, production environment in \S\ref{sec:detector}.

\subsection{Geographical Routes of Phishing Emails}
\label{sec:sendercountry}
\enew{In this section, we examine the geographical distribution of the email delivery infrastructure used to send phishing emails.} 
We matched the IP addresses across our dataset's \textsc{Received} headers to their country using the MaxMind Geolite2 database~\cite{maxmind}. We matched addresses that MaxMind did not geolocate by using WhoIS to identify the country that the IP address was registered under.

There are some caveats with this analysis due to the inaccuracy of geolocating IP addresses. For example, prior work notes that, while geolocation databases obtain good accuracy (85\%, and MaxMind reports a coverage of 80\%), they are not perfect, with the two leading factors behind errors in geolocation are IP addresses that belong to ASes with a global presence and IP addresses that change ownership through merger and acquisition~\cite{geolocateacc}. Our analysis attempts to address some of these limitations by studying location at the country granularity, which MaxMind claims are 99.8\% correct~\cite{maxmind} \enew{ and which prior work shows to be reasonably accurate~\cite{geolocationCountries}}, but we acknowledge that imperfect data might inherently lead to some inaccuracies. Furthermore, results can vary based on a company's location and the location of those who send it legitimate email. Our results are specific to the \cuda customers in our dataset, which are largely based in the US.

\begin{figure}
    \centering
    \includegraphics[width=0.6\columnwidth]{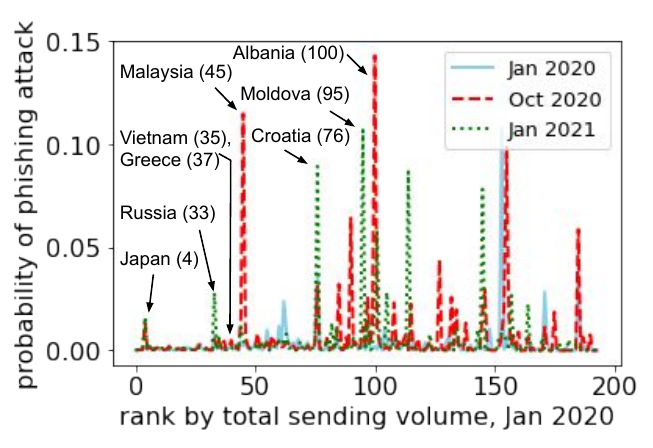}
    \caption{Probability of sending phishing for countries that sent over 1,000 emails across 3 months, ranked by volume in Jan '20.}
    \label{fig:probsender}
    \vspace{-3ex}
\end{figure}

\paragraph{Probability of phishing by origin country:}
For countries with more than 1000 emails in our dataset, \Cref{fig:probsender} shows the probability of emails from each country that were phishing. 
Similar to the distribution by AS, the countries with the highest email volume (by originating IP address) also have a low probability of phishing. However, several countries have a significant amount of outbound email that contains a high proportion of phishing emails.
For example, in at least one month of our dataset, five countries rank in the top-100 countries by sending volume where over 5\% of these emails are phishing.

\paragraph{Geographic routes of phishing emails:}

\begin{figure}[t]
\begin{minipage}[t]{0.5\linewidth}
    \centering
    \includegraphics[width=\columnwidth]{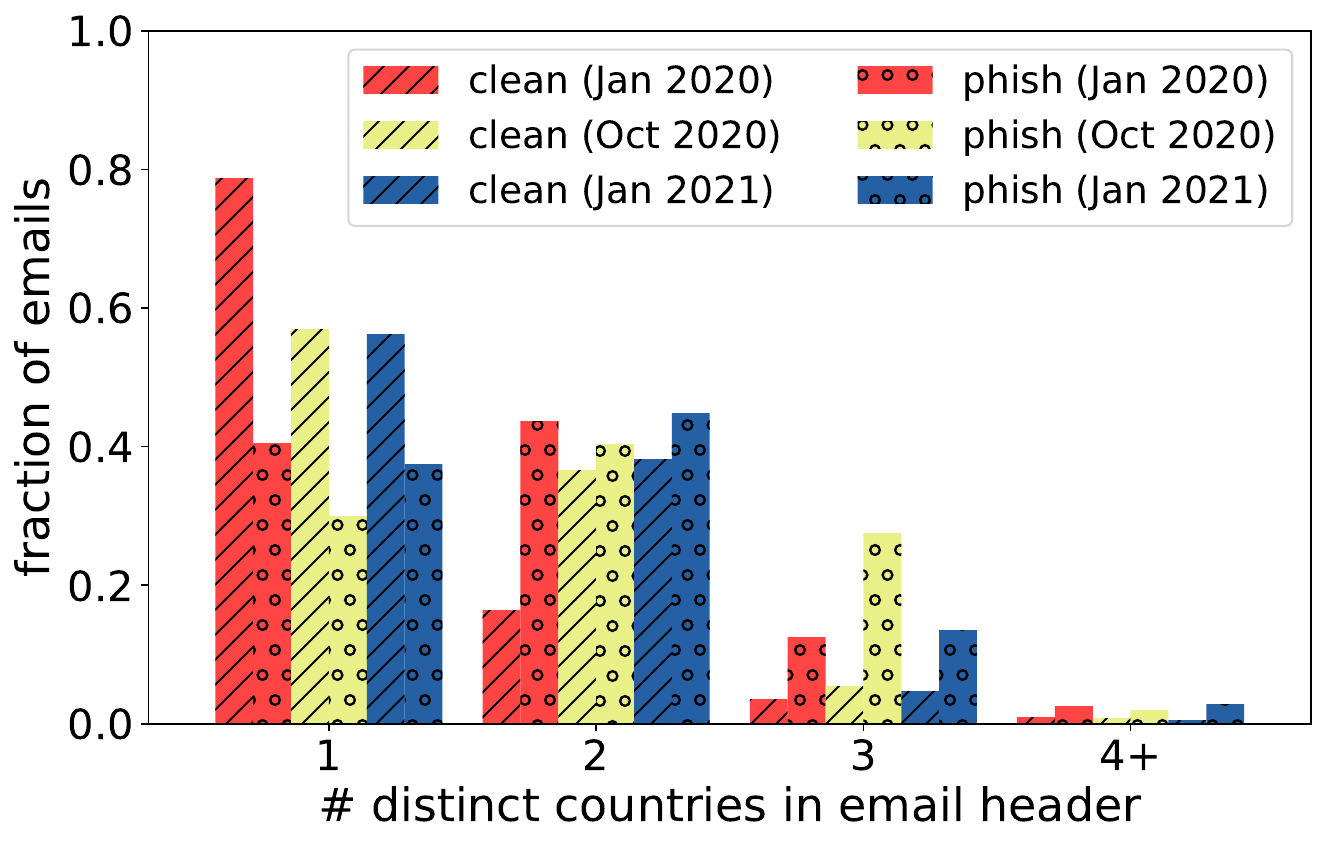}
    \caption{The distribution of countries in an email's relay path across our datasets for clean and phishing emails.}
    \label{fig:distinctcountry}
    \end{minipage}
\hspace{0.5cm}
\begin{minipage}[t]{0.5\linewidth} 
    \centering
    \includegraphics[width=\columnwidth]{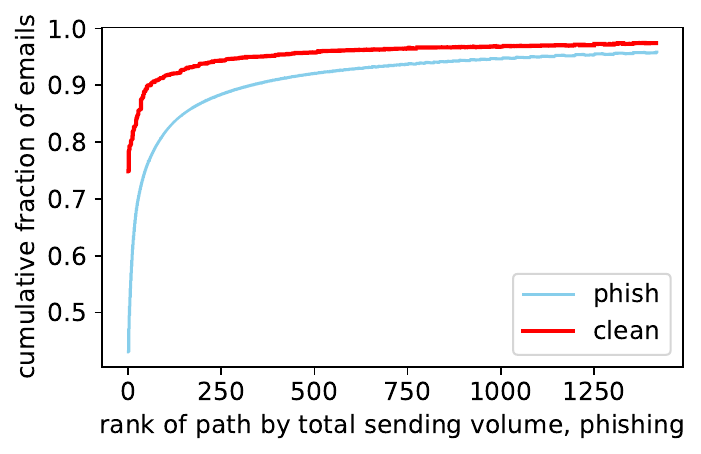}
    \caption{Cumulative fraction of emails that traverse specific country-routes ranked by phishing volume per country-route.}
    \label{fig:routes}
    \end{minipage}
    \vspace{-1.5ex}
\end{figure}
Widening our geographic analysis from just the origin IP address to the full email relay path, 
we found that phishing emails tend to route through more countries than clean emails (Figure~\ref{fig:distinctcountry}). In terms of exact paths, most emails, clean and phishing, route through a small number of distinct country sequences. Figure~\ref{fig:routes} shows that after mapping the IP addresses in an email's delivery path to their country, the top-25 country routes account for over 60\% of phishing traffic in our dataset. 
The most common path taken by both clean and phishing emails traversed servers that resided solely within the US, reflecting the large proportion of US-based organizations in our data.
However, we find that this country route accounts for over 70\% of traffic for all clean emails across all time periods in our data, while less than half of all phishing attacks were routed only through the United States. The second-most-common route was traffic routed from US-based servers to UK-based servers, while the third-most-common route was from Germany to the US. Both of these paths routed a slightly higher fraction of the overall phishing emails than their fraction of clean emails, but these routes each carried 5\% or less of the total emails each month. 

\paragraph{Takeaways:}
\enew{Our analysis shows that the number of countries in the email delivery path and the countries themselves within the path could provide a useful feature to improve phishing detection, as long as a defender accounts for limitations of geolocation and dataset biases.} 

%% file: case_studies.tex
\section{Case Studies}
\label{sec:caseStudies}
\enew{In this section, we investigate several nuances in network behavior 
by examining some illustrative examples of ASes (and the IP addresses contained within them) in both the low and high-phishing concentration categories. First, we take a deeper dive into the two prominent low-concentration networks, Amazon and Microsoft~(\S\ref{sec:case:low_risk}). 
Next, we explore two contrasting examples of high-phishing-concentration networks: one that remains high concentration across the three months~(\S\ref{sec:case-study-persistent}), and one that is comparatively transient, sending phishing emails in short bursts in a single month~(\S\ref{sec:case-study-bursty}).} In particular, we find that many networks are quite unstable (over even a period of one month) with respect to their phishing email traffic, motivating the need for more dynamic solutions than static blocklists.

\subsection{Low Phishing Concentration Networks}\label{sec:case:low_risk}
\enew{We examine specific IP addresses in ranges owned by Amazon and Microsoft to better understand why a significant fraction (31\%) of all phishing email in our dataset originate from ASes owned by these two companies (however, phishing still constitutes less than 0.01\% of the total email sent from these networks).}
In some cases, we find that attackers use hosting services provided by Amazon and Microsoft (i.e., AWS and Azure) to send phishing email (whether it be through compromising or deploying a mail server on these platforms). Additionally, we detect a potential case of account hijacking.



\begin{figure*}[t]
    \centering
     \begin{subfigure}{0.45\textwidth}
         \centering
         \includegraphics[width=\textwidth]{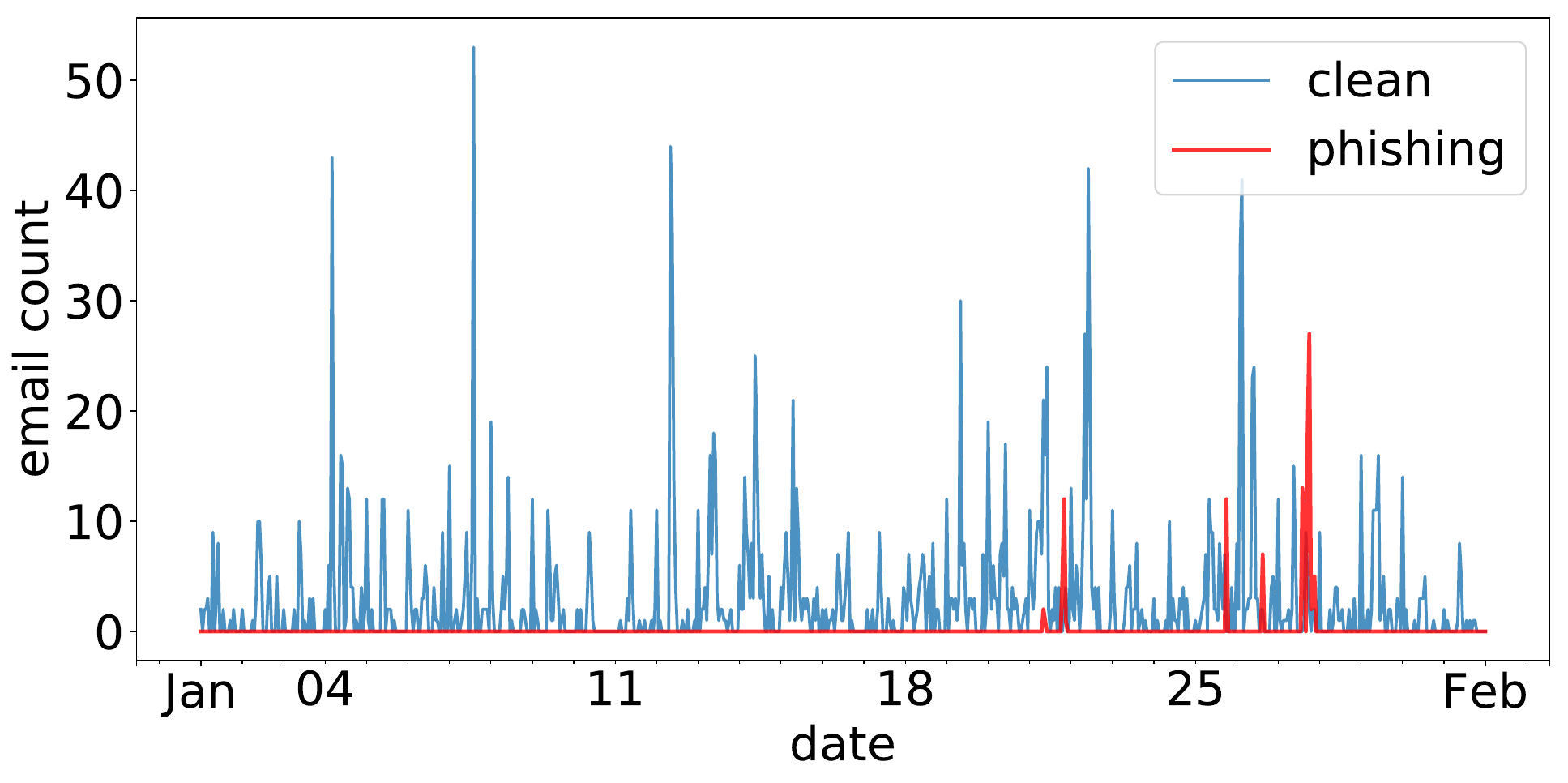}
        \subcaption{Suspected compromised email account from Amazon AS 16509.}
         \label{fig:accnt}
     \end{subfigure}\hspace{3ex}
     \begin{subfigure}{0.45\textwidth}
         \centering
         \includegraphics[width=\textwidth]{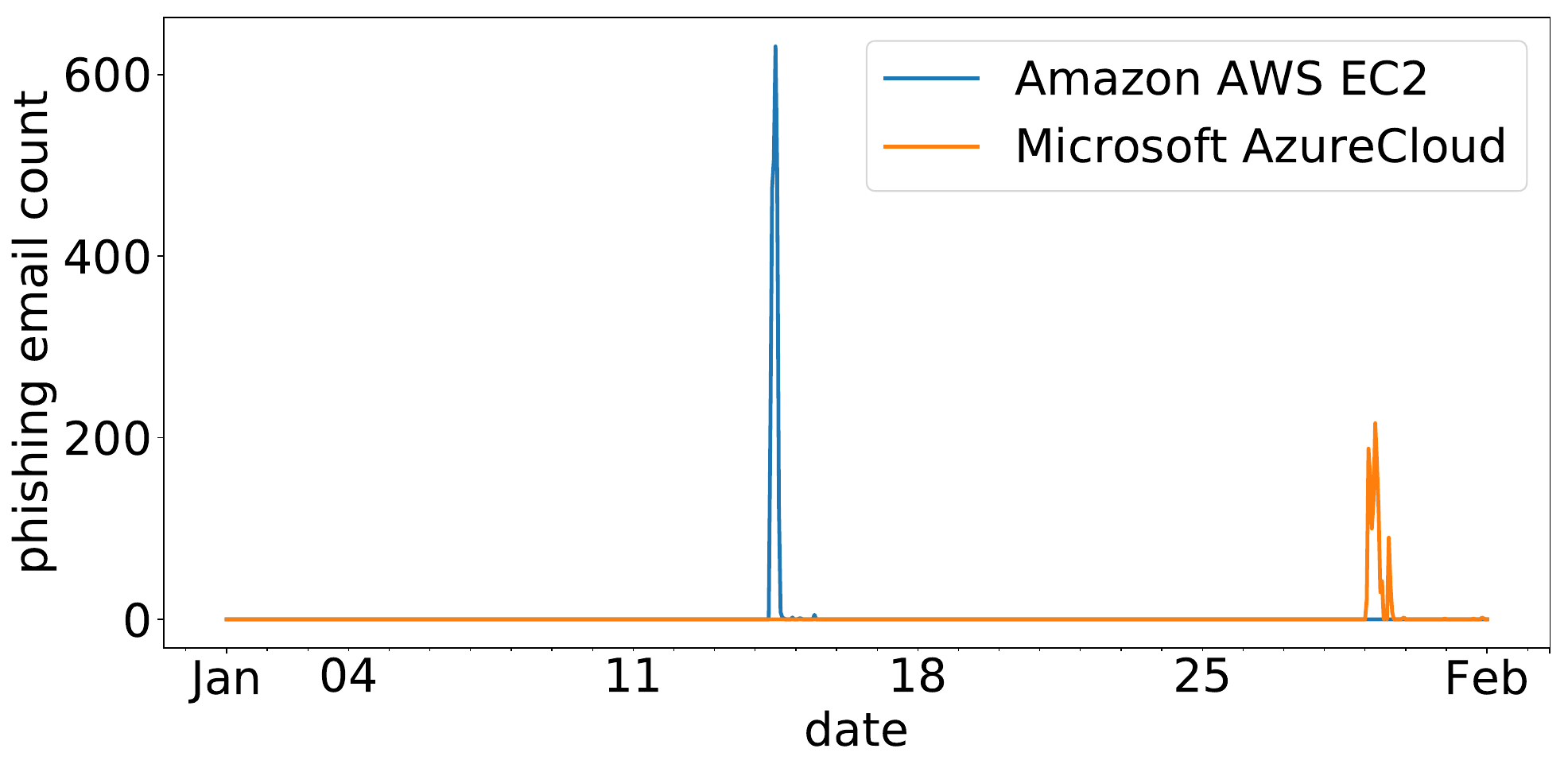}
        \subcaption{Highest-volume phishing sender from Amazon AS 16509 and Microsoft AS 8075.}
         \label{fig:amazonAndMicrosoftSending}
     \end{subfigure}
     \begin{subfigure}{0.45\textwidth}
         \centering
         \includegraphics[width=\textwidth]{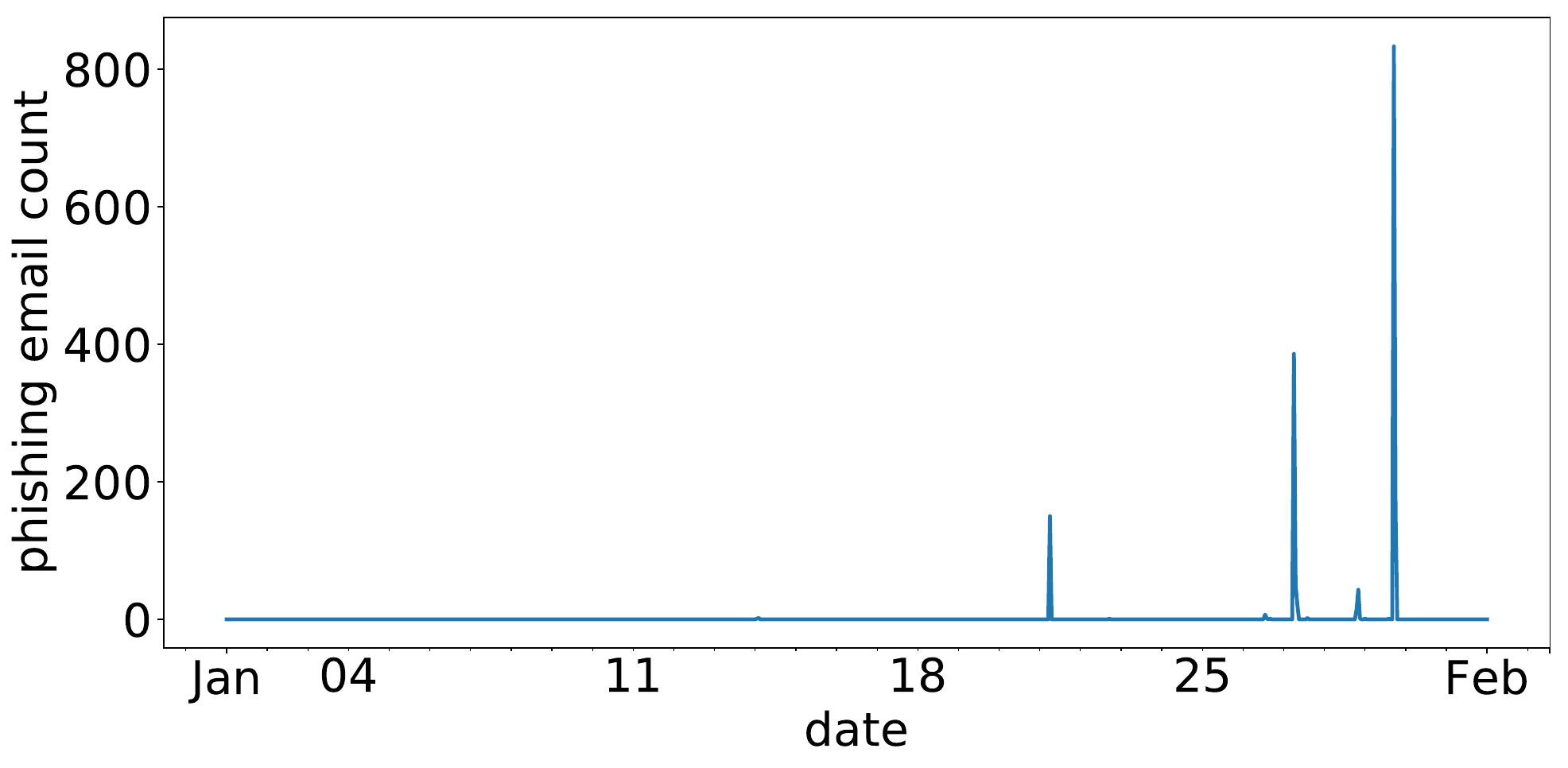}
         \subcaption{Phishing emails from AS 52000.}
         \label{fig:burst}
     \end{subfigure}\hspace{3ex}
     \begin{subfigure}{0.45\textwidth}
         \centering
         \includegraphics[width=\textwidth]{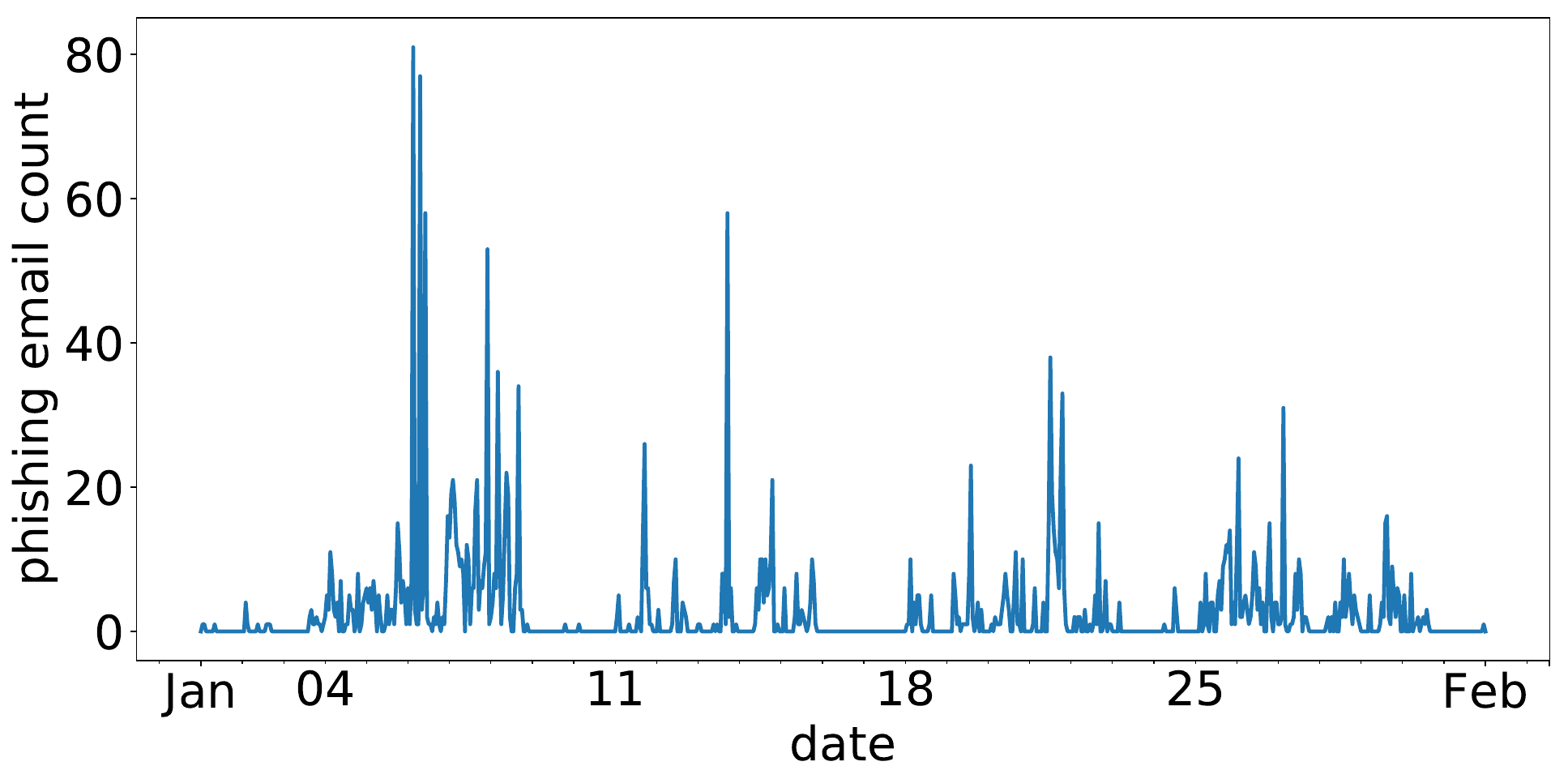}
         \subcaption{Phishing emails  from AS 9009.}
         \label{fig:m247}
     \end{subfigure}
     \caption{Phishing over time for particular IP addresses and ASes during January 2021.}
     \vspace{-3ex}
\end{figure*}

\vspace{-1ex}
\subsubsection{Amazon: AS 14618 \& AS 16509}\label{sec:amazon}
Based on the aggregate email traffic from the Amazon IP address responsible for the largest number of phishing email (1,324 emails in January 2021) and additional information from the email headers, we identify this IP address, which belongs to a published EC2 IP address range\footnote{On aggregate, 39.36\% (7,623 emails) of all phishing email originating from Amazon's ASes come from a published EC2 IP address range (based on Amazon's published IP address ranges, we were unable to determine the other Amazon services for the remaining phishing emails)~\cite{AWSrange}.}, to correspond to a ``dedicated" phishing email server. The emails originating from this IP address contained 388 unique \from values where the email address was \textit{noreply$@$domain.com} across different domains. This information suggests that the attackers were spoofing the \from value in their attacks. Furthermore, all the phishing emails from this IP address were sent within a short period of time (as shown in the blue curve in Figure~\ref{fig:amazonAndMicrosoftSending}). We captured 458 emails that were not classified as phishing from this IP address, and did not see any phishing emails in other time snapshots. This large burst of seemingly spoofed emails suggest the attackers either set up their own email server on AWS or compromised a legitimate server to send a large amount of phishing emails.

We were also able to identify a phishing attack potentially enabled by account hijacking. We found a smaller volume of phishing email originating from a single Amazon IP address, where the \from address contained a legitimate corporate domain and a human-looking user name (\eg \textit{jane.smith$@$company.com}). The account sent 115 phishing emails to \cuda customers in January 2021, but did not send any phishing emails in January 2020 or October 2020. Moreover, we found about 1,500-2,000 clean emails originating from this IP address in our dataset. Given the low proportion of phishing email and longer duration of approximately 1 week (as shown in Figure~\ref{fig:accnt}), which matches the findings of prior work on compromised email accounts that shows that compromise incidents can last for days and even weeks~\cite{lateral1,compAccoountsKdd2010}), we find it likely that this is a case of email account compromise. This case may benefit from a separate defense, such as looking for unusual patterns in account activity~\cite{compAccoountsKdd2010}.

\vspace{-1ex}
\subsubsection{Microsoft: AS 8075}\label{sec:microsoft} 
\enew{Similar to the first case we studied in Amazon's ASes, we observe that the IP address in Microsoft's AS that sent the most phishing emails behaves most like an attacker-controlled server. In total, this IP address, which falls within an Azure Cloud range\footnote{Among the phishing emails from Microsoft's networks, we determined that 99.6\% came from Azure Cloud IP address ranges, with the rest from Exchange Online and Sharepoint~\cite{Azurerange,Microsoft365range}.}, sent 2,465 phishing emails and 20 clean emails in January 2021, with 578 unique \from values, where the \from values for all the phishing emails are in the form of \textsc{noreply$@$domain} with different domains. Given the high number of emails, the wide-range of unique domains, and the generic \textsc{noreply$@$} addresses, this scenario is most consistent with an attacker using email spoofing to forge sender names. This IP address also exhibited a ``bursty'' email sending behavior (as shown in the orange curve in Figure~\ref{fig:amazonAndMicrosoftSending}), similar to the first Amazon case study (the blue curve), and we did not find any phishing email from this IP address in the 2020 datasets.}

The fact that the highest volume servers send their attacks in such short bursts~(Figure~\ref{fig:amazonAndMicrosoftSending}) suggests that they may be shut down after a short time \textit{and} that the pattern may be exploited as part of detection. Cloud providers  could use these patterns as a signal to identify VMs abused for sending malicious email. Similarly, enterprise email defenses can potentially quarantine large email bursts from a previously-unseen sending server. \enew{This attack strategy also illustrates the limitations of using purely network-reputation-based defenses against phishing.}



\subsection{High Phishing Concentration Networks}
\label{sec:caseStudy_highRisk}
\enew{In \S\ref{sec:temporal} we found that some high phishing concentration networks remain consistently classified as such over our three time periods, while others are transient. Here, we examine each of the two cases: one that sends phishing emails during only one month in our dataset (and displayed similar patterns to our Amazon and Microsoft case studies), and one that consistently sends many phishing emails over the three time periods. In particular, we highlight the instability (``burstiness") of the IP addresses within these ASes responsible for sending phishing email.}

\begin{table}[t]
\begin{minipage}[t]{0.5\linewidth}
\centering
\resizebox{.9\textwidth}{!}{
\begin{tabular}{|l|l|l|}
\hline
\textbf{Phishing Lifespan} &
  \textbf{\# Phishing} &
  \textbf{\# Clean} \\ \hline
15 days 11:59:34 & 245 & 10 \\ \hline
1 days 16:32:37  & 127 & 19 \\ \hline
0 days 09:10:08  & 96  & 2  \\ \hline
10 days 12:38:23 & 81  & 0  \\ \hline
9 days 08:24:46  & 71  & 2  \\ \hline
1 days 17:15:40  & 71  & 0  \\ \hline
7 days 23:46:46  & 67  & 88 \\ \hline
21 days 03:07:51 & 64  & 4  \\ \hline
0 days 05:47:59  & 60  & 3  \\ \hline
0 days 01:03:14  & 53  & 0  \\ \hline
\end{tabular}%
}
\caption{Phishing sending lifespan and email volume for the top-10 phishing senders by IP within AS 9009 in January 2021.}
\label{fig:top10-9009}
\end{minipage}
\begin{minipage}[t]{0.5\linewidth}
\centering
\begin{tabular}{|l|l|l|}
\hline
\textbf{Time Period} & \textbf{\# Phishing} & \textbf{\# Clean} \\ \hline
Jan 2020 & 955  & 29099 \\ \hline
Oct 2020 & 3207 & 47560 \\ \hline
Jan 2021 & 2135 & 8685  \\ \hline
\end{tabular}
\caption{Clean and phishing email volume over time of AS 9009.}
\label{tab:persistent-case-study}
\end{minipage}
\vspace{-2em}
\end{table}

\vspace{-2ex}
\subsubsection{Bursty High Phishing Concentration Network: AS 52000}
\label{sec:case-study-bursty}
Attackers used the IP addresses from AS 52000 (a Russian hosting company with data centers in the Netherlands, Russia, and the U.S.) to send phishing email for very short lifetimes: across the entire month, we see phishing email from these addresses for less than 24 hours total, and the highest volume phishing email senders among these IP addresses sent all of their attacks within short 30-minute windows.
We saw two large bursts of phishing email in January 2021 (Figure~\ref{fig:burst}) comprising a total of 1,673 phishing emails from AS 52000; no other emails in our dataset originate from this network during that month. We found phishing emails from 20 unique IP addresses within this AS, and the emails span 533 campaigns (as defined in~\S\ref{sec:dataset-desc}). No phishing emails originated from this AS outside of these two bursts in Jan 2021. 

This behavior suggests an attacker who abuses cloud-hosted servers to send large bursts of phishing emails before switching to new infrastructure (\eg because a provider terminates their services). While on aggregate AS 52000 (and the IP addresses within it) have a poor reputation (in terms of the proportion of fraudulent activity detected), the burstiness of the phishing email traffic means that these IP addresses would be unlikely to end up on established blocklists. Instead, a time-series-based anomaly detector that analyzes the volume of a particular network's email traffic could potentially flag periods of higher phishing risk, which smaller hosting companies could employ to flag suspect servers.

\vspace{-1ex}
\subsubsection{Persistent High Phishing Concentration Network: AS 9009}\label{sec:case-study-persistent}
We observed a consistently high volume and proportion of phishing email originating from AS 9009 (which belongs to a UK-based hosting company). This AS was one of the 9 networks we classified as high phishing concentration for all three time periods (Table~\ref{tab:persistent-case-study}). In contrast to the bursty behavior observed in the prior case studies, as shown in Figure~\ref{fig:m247}, we find that phishing attacks regularly originated from this network throughout the month of January 2021 (the other months in our data had similar distributions). 
The phishing emails from AS 9009 span 1,031 unique phishing campaigns and come from 151 unique IP addresses, where 32 of these IP addresses send over 20 phishing emails each. 
One reason for the high number of unique phishing IP addresses and campaigns might be that multiple attackers use this AS to send phishing emails.

To shed light on the volatility of the phishing IP addresses within AS 9009, we calculated the phishing email sending lifespan of each IP address, \ie the time between the first phishing email we saw from the IP address to the last phishing email (because we compute this duration over 1 month windows, the maximum lifespan is 31 days). In January 2021, the average lifespan of a phishing IP address from this AS is 41.5 hours. However, as shown in Table~\ref{fig:top10-9009}, many of the IP addresses with the highest phishing volumes continuously send phishing emails for extended periods (weeks) of time. In this case, IP-based reputation features would be useful for blocking phishing email. Furthermore, this longer-lived malicious behavior suggests the need to adopt additional technical or policy-based defenses to help curtail attacks coming from such networks.

%% file: detector.tex
\section{Experiences In Production}
\label{sec:detector}
\enew{In Sections~\ref{sec:temporal} and~\ref{sec:case:low_risk}, we showed that the set of networks sending phishing emails is not stable over time.
Many networks only send high volumes of phishing emails for a limited time (4.3), potentially reflecting the behavior of account takeover (5.1.1) or compromised Azure or EC2 instances (5.1.2). 
Because this analysis suggests that phishing detection rates can be improved by accounting for these temporal conditions, we worked with \cudaFullName to deploy a classifier that better adapts to changing attacker behavior. We found that our approach enables \cuda to identify phishing attacks that were not identified by their existing email classifiers or by pre-filters that customers already deployed, without incurring additional false positives.}

\begin{figure}[t]
    \centering
    \includegraphics[width=0.6\columnwidth]{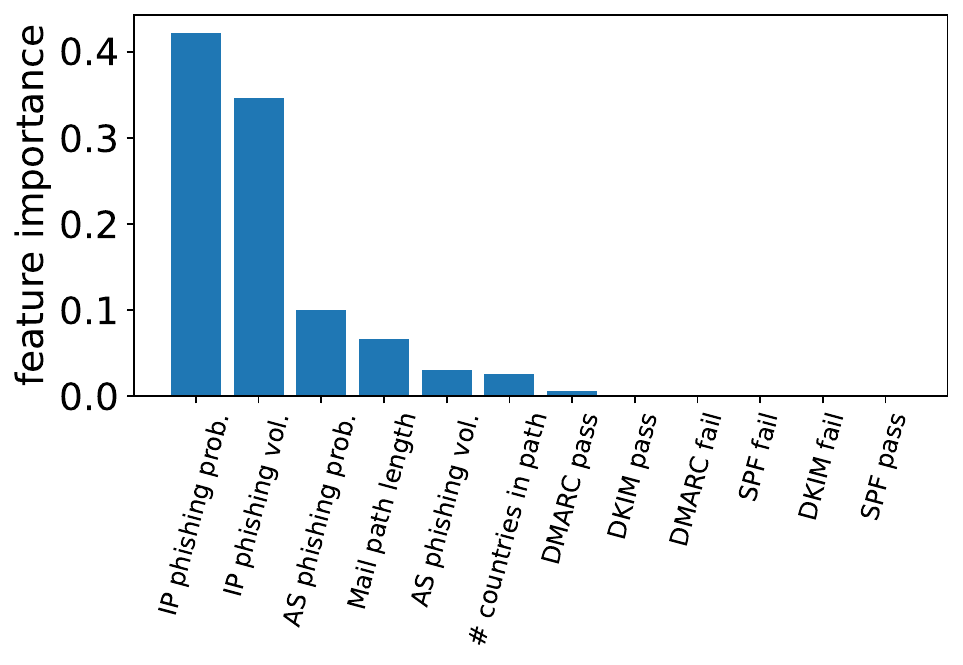}
    \caption{Relative importance of features.}
    \vspace{-.7em}
    \label{fig:featurerank}
\end{figure}

\paragraph{Feature importance for phishing detection:}\
First, as a proof of concept and
to better understand which features are more important for phishing detection, we trained a Random Forest model on 12 features based on our analysis in \S\ref{sec:results} to differentiate clean and phishing emails in our dataset. Figure~\ref{fig:featurerank} depicts the relative importance of these features based on Gini importance~\cite{gini}.
The proof of concept classifier confirmed our understanding that probability and volume of phishing by origin IP address were indeed the most important for phishing detection. Because of the complexities and real-world impact of modifying production classifiers, we chose to use only these two features (which were not already incorporated in any of Barracuda's existing classifiers).

\paragraph{Production classifier design:} Our classifier aims to better adapt to networks that only send high volumes of enterprise phishing emails for a relatively short period of time, while still capturing persistently malicious senders. To this end, we designed our classifier to dynamically recalculate, on a daily basis, the features it uses to make a decision about whether an email is phishing or not. To do this, for each IP address from which \cuda sees emails, it counts the number of phishing emails and the number of clean emails (according to \cuda's existing classifiers) seen in a sliding window of the last $n$ days. While some high-phishing-concentration IP addresses are only active for 30 minutes (e.g., \S\ref{sec:case-study-bursty}), we found through testing different time windows that the day-granularity best captured phishing email while minimizing false positives. 
Then, our detector uses the number of phishing emails and the ratio of phishing to clean emails to calculate a ``phishing-risk" score. 
This score is used to make an independent, binary decision about whether an email that has previously been classified as benign by any of \cuda's production classifiers is phishing. 
We first conducted offline experiments with historical phishing detection data, testing the performance of window sizes $n$ from 0-90 days and different thresholds for the ``phishing-risk" score. Then, to determine their optimal values in \cuda's production environment, we conducted smaller-scale online experiments with some promising values derived from the offline experiments.
Importantly, any attacks our classifier finds were \textit{previously undetected} by \cuda's existing classifiers.

\paragraph{Classification Results:} \enew{Over a span of nearly 5 months, the new detector increased the number of emails flagged as phishing by 3-5\% per day (beyond those detected by \cuda's existing detectors), with no false positives based on manual inspection. For one day per week during the first month of deployment, members of \cuda’s security team manually examined all emails that were classified as attacks by the new detector, or reported by their users. Interestingly, manual inspection found that this detector detected phishing and also other types of email attacks (e.g., business email compromise, scams, and spam).}

\paragraph{Takeaways:}
\enew{In a real-world deployment, we demonstrated that capturing the changes of networks' email sending behavior over time can lead to an improved phishing detection rate, without an increase in false positives. In particular, our classifier demonstrates a strict improvement over static sender-reputation features - we are able to reliably find additional attacks even though some spam filters used by organizations in our dataset already employ sender reputation scoring heuristics~\cite{autowhitelist}. We note that an attacker could evade this defense by sending a low volume of phishing emails from a large number of IP addresses, but such a strategy would incur additional operational costs and could still be (partially) blocked by existing methods.}

%% file: discussion.tex
\section{Conclusions and Key Lessons}
\label{sec:discussion}

We presented a large-scale analysis of the characteristics of the network delivery infrastructure behind phishing emails targeting enterprises.
Our findings
provide a useful avenue for improving phishing defenses, which we demonstrate apply in a real-world production environment with the classifier we deployed at \cuda that was able to block an additional 3-5\% of previously undetected phishing emails.
We distill takeaways for how the community can apply both technical and policy-based defenses based on our results. 

\paragraph{Features targeting the network origins of an email are promising:}
Our analysis revealed that most phishing emails in our dataset came from a few hundred IP addresses and ASes.
Additionally, excluding Amazon and Microsoft, the top-100 ASes by phishing volume account for fewer than 5\% of clean emails in our dataset, and many phishing emails come from what we termed \emph{high-concentration networks}, those with more than 2\% of their emails labeled as phishing~(\S\ref{sec:infrastructure}). 
Given that these networks are both the origins of a substantial volume of phishing emails and have a relatively large ratio of phishing to clean emails, reputation features about an email's originating network have the potential to provide high-impact signals of phishing. \enew{However, since a non-negligible volume of clean email originates from these same networks, strict blocklists may yield a higher-than-acceptable false positive rate. Instead, our results suggest that combining these features with others will improve phishing email detection while maintaining a low false positive rate.}

\paragraph{Network phishing behavior over time is an important consideration for detection:}
\enew{Our results showed that some networks have highly variable amounts and/or proportions of phishing emails that they send over time (\S\ref{sec:temporal}). This phenomenon provides a possible explanation for why static lists of suspicious sender IP addresses or ASes, such as those used by many organizations in our dataset,
prove insufficient at defending against enterprise phishing attacks.
Our findings suggest that defenders need more agile and flexible methods of incorporating network-based characteristics, rather than static approaches like blocklists, to successfully defend against phishing attacks.
To illustrate this, we showed that, in a production environment, incorporating dynamically updated features can help uncover a significant number of previously-undetected phishing emails~(\S\ref{sec:detector}).}

\paragraph{Internet hosting services are a major source of phishing:}
The rise of Infrastructure-as-a-Service (IaaS) has provided an easy path for attackers to acquire infrastructure (\cref{sec:infrastructure} and~\cref{sec:caseStudies}), with
servers on prominent and reputable cloud hosting providers, such as Amazon AWS and Microsoft Azure, 
being responsible for nearly one-third of all phishing emails in our dataset. An additional quarter of the phishing emails come from smaller cloud hosting companies that are among the \emph{high-concentration networks} we identified.
These results suggests that Internet hosting companies are well-positioned to significantly curtail the volume of phishing emails.
Future work should investigate whether networks operators can take steps to detect and stop \textit{outbound} emails sent from their infrastructure, 
or whether adopting stricter security policies around email originating for cloud hosting services can decrease their phishing volume. 

\paragraph{Attackers face insufficient defensive pressure on their delivery infrastructure:}
Many networks consistently send a high volume or concentration of phishing emails across multiple months in our datasets~(\cref{sec:temporal}), suggesting that attackers do not face significant pressure to change servers and that information about an email's delivery origin remains an underutilized avenue for mitigating phishing attacks. We hope our results provide impetus for changing this.

%% file: appendix.tex
\appendix

\section{Ethical Considerations}
\label{appendix:ethics}
In addition to consulting with and obtaining approval from our institution's IRB (AAAT8774 (Y01M00)), we carefully considered the ethics of our research along three key dimensions discussed in the Menlo Report: 
(1) does our research present minimal risk to the well-being and rights of participants?
(2) does our work take appropriate measures to minimize such risks?
(3) do the benefits of our research outweigh the risks?

In terms of benefits from the research, improving the community's defenses and understanding of phishing attacks has clear benefits both to the organizations in our study and to society at-large, given the widespread threat posed by enterprise phishing attacks.
Soundly and accurately studying the problem of enterprise phishing fundamentally requires analysis of large-scale, real-world email data, which makes accessing and using our study's data inherently necessary.
To minimize the harms of using such data, the clean emails that we have access to in our dataset (i.e., emails from legitimate, non-attacker users), do not include sensitive attributes such as the email message body, subject line, or sender name and email address.
The data in our study comes from enterprise email accounts, whose organizations have consented to using their data for research purposes such as ours. Given the scale and nature our our dataset, obtaining individual consent from each user is clearly infeasible.
Furthermore, the email accounts involved in our dataset are owned and provided by each enterprise, and it is commonplace and reasonable for enterprises to monitor and scan their employees' work account emails for malicious activity.
As such, our work's use of this data does not increase the risks or jeopardize the well-being of these users, particularly given that our analysis involves a strictly less sensitive version of this already-collected data.
Weighing this collection of factors, such as the minimal risk to participants' welfare posed by our study, against the benefits of our research, we believe our work properly aligns with the ethical norms and principles of our field. 

\section{Email Network Origin Validation}
\label{appendix:spoofing-validation}

We identify cases that are not forged based on the following four tests using the IP address of an email's MX-identified sender and purported origin IP address (the first public and non-reserved IP listed in the \textsc{Received} headers).
Combining tests, we find at least 90.6\% of phishing emails in our dataset do \textit{not} spoof the origin IP addresses in their \textsc{Received} headers, at least in ways that impact our analysis.

First, for 66\% of phishing emails, the IP address of the MX-identified sender and the origin IP address belong to the same AS.
Since most of our analysis focuses on the AS-granularity, this criteria ensures that any spoofing for this set of phishing emails will not distort or bias our results.
To calibrate expectations, we ran the same analysis on the benign emails in our dataset, for which the sender has no reason to forge their headers; 73\% of these emails had the same AS for both their purported origin IP address and the IP address of the MX-identified sender (a similar proportion to the phishing emails). 
Second, if a phishing email's pair of MX-identified sender IP address and origin IP address exactly matches a pair seen for benign emails, then we also label these as instances where the \textsc{Received} headers were not modified. In total, an additional 16.2\% of the total phishing emails matched this criteria.
Third, an additional 2\% of the total phishing emails satisfied an even stronger property, where their entire relay path (all IP addresses in their \textsc{Received} header) had an exact match with a path seen in our benign email dataset.
Finally, for an additional 6.4\% of phishing emails, the IP address listed in the recipient's MX record only appeared once in the \textsc{Received} headers, and it matched the purported origin IP address in the email exactly (5.9\%) or belonged to the same AS (0.5\%). These appear to be internal emails that were flagged by O365 as emails sent from an external source. Such cases can occur when emails are sent between different domains that belong to the same O365 account. Such single AS paths are not the focus of our paper's analysis, which investigates the AS or country-level characteristics of the delivery path before the recipient's servers.

\begin{figure}
    \centering
    \includegraphics[width=0.6\columnwidth]{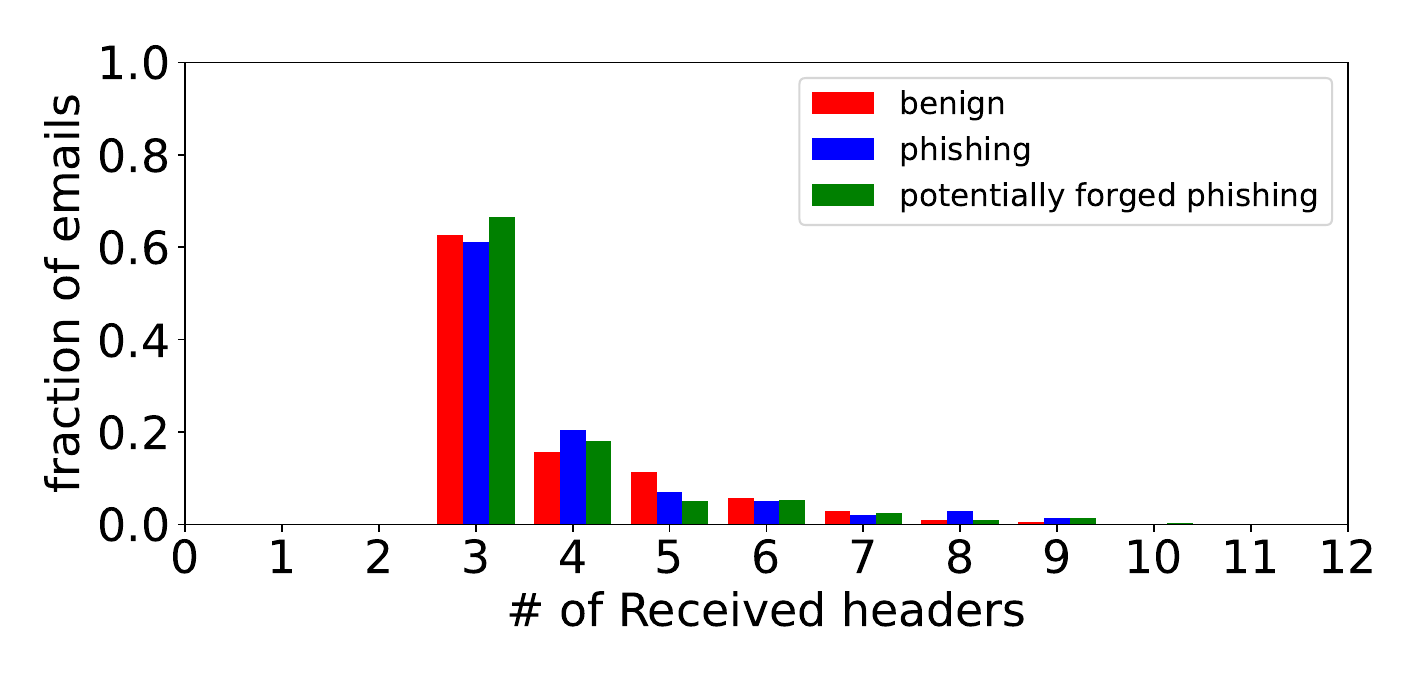}
    \caption{Distribution of path lengths for emails with $\geq$ 3 headers. The green bars show the distribution for the 10\% of phishing emails that do not match one of our validation criteria (\S\ref{sec:spoofing}).}
    \label{fig:chainlen}
    \vspace{-3ex}
\end{figure}

The set of heuristics above cover  90.6\% of all of the phishing emails in our dataset.
For the remaining 9.4\%, we investigate the distribution of path lengths (total number of \textsc{Received} headers) and compare to the path lengths of the clean emails in our dataset. The remaining 9.4\% includes only emails with at least three headers, since empirically, our tests above established that all shorter paths were not forged. So, we compare the distribution of path lengths of these remaining emails to clean and phishing emails with at least three headers.  
As seen in Figure~\ref{fig:chainlen}, the path length distribution between the remaining phishing emails (green) is very similar to the clean emails (and to the overall set of phishing emails) in our dataset.

This similarity suggests that attacker manipulation did not have a significant impact on the paths we observe,  
since such tampering would lead to a longer path length (assuming that the attacker is positioned at the beginning of the email delivery path).

\clearpage